\documentclass[12pt]{article}
\usepackage[margin=1in]{geometry}
\usepackage{amssymb,amsthm}
\usepackage{bbm}
\usepackage{bm}
\usepackage{upgreek}
\usepackage{multirow}
\usepackage{amsmath}
\usepackage{graphicx}
\usepackage{natbib}
\usepackage{subcaption}
\usepackage{algorithm}
\usepackage{algorithmic}
\usepackage{url}
\usepackage{tikz}
\usepackage[T1]{fontenc}
\usepackage[utf8]{inputenc}
\usepackage{authblk}
\usetikzlibrary{positioning,chains,fit,shapes,calc,arrows}
\newcounter{dummy}
\usepackage{enumitem}
\makeatletter
\newcommand\myitem[1][]{\item[#1]\refstepcounter{dummy}\def\@currentlabel{#1}}
\makeatother
\newcommand{\Var}{\operatorname{Var}}
\allowdisplaybreaks
\tikzstyle{int}=[draw=none,  minimum size = 2em]
\tikzstyle{init}=[pin edge={->,thin, black}]

\title{Sequential Estimation of Temporally Evolving Latent Space Network Models}

\author[1]{Kathryn Turnbull\thanks{k.turnbull@lancaster.ac.uk}}
\author[1]{Christopher Nemeth}
\author[2]{Matthew Nunes}
\author[3]{Tyler McCormick}
\affil[1]{Department of Mathematics and Statistics, Lancaster University}
\affil[2]{School of Mathematical Sciences, University of Bath}
\affil[3]{Department of Statistics, University of Washington}

\date{}

\begin{document}

\maketitle

\begin{abstract}
  In this article we focus on dynamic network data which describe interactions among a fixed population through time. We model this data using the latent space framework, in which the probability of a connection forming is expressed as a function of low-dimensional latent coordinates associated with the nodes, and consider sequential estimation of model parameters via Sequential Monte Carlo (SMC) methods. In this setting, SMC is a natural candidate for estimation which offers greater scalability than existing approaches commonly considered in the literature, allows for estimates to be conveniently updated given additional observations and facilitates both online and offline inference. We present a novel approach to sequentially infer parameters of dynamic latent space network models by building on techniques from the high-dimensional SMC literature. Furthermore, we examine the scalability and performance of our approach via simulation, demonstrate the flexibility of our approach to model variants and analyse a real-world dataset describing classroom contacts. 
\end{abstract}


\section{Introduction}
\label{sec:intro}

Dynamic network data describe the evolution of interactions among a population of interest over time. Data of this type arise in a diverse range of applications and may represent, for instance, friendships within a social group \citep{krivitsky2014}, coauthorship among academics \citep{yang2011} or co-evolution of financial indices \citep{durante2014}. A distinction can be made between settings in which the population remains fixed (see \cite{sarkar2006, xing2010, durante2016lady}) or evolves (see \cite{barabasi1999, bloemreddy2018}) over time and, in this article, we focus on the former setting. By incorporating the temporal aspect of network data, we are able to analyse how interaction patterns evolve over time and make predictions about future connections.

Several temporal extensions of popular models for static network data have been proposed in the literature (for example, see \cite{kim2017, hanneke2010, krivitsky2014, xing2010, zhang2017}) and we focus our attention on the latent space framework of \cite{hoff2002}. In this approach, each node is associated with a low-dimensional latent coordinate which controls the tendency for ties to form between nodes, and it is typical to assume that nodes whose latent coordinates are close in terms of Euclidean distance are more likely to be connected. This formulation allows a practitioner to determine an intuitive visualisation of the data, and encourages transitive relationships since the Euclidean metric satisfies the triangle inequality. Dynamic model variants of the static models proposed in \cite{hoff2002} have been developed for both discrete time (see \cite{sarkar2006, sewell2015, friel2016}) and continuous time (see \cite{durante2014, rastelli2021}) and, in the discrete time case, it is common for authors to rely on a state space model (SSM) formulation in which the latent trajectories evolve according to a Markov process.

In this article, we consider the novel application of Sequential Monte Carlo (SMC) methods with dynamic latent space network (DLSN) models. Posterior samples for parameters of a DLSN model are typically obtained via MCMC \citep{sewell2015, durante2014} which allows for asymptotically exact inference. However, standard algorithms become increasingly computationally costly as the dimension of the state space increases and, in this setting, this can be due to the number of nodes in the network or the number of observations in time. To improve scalability, several authors have proposed likelihood approximations \citep{raftery2012, rastelli2018} whereas others rely on variational inference which targets a computationally cheaper approximation of the posterior (see \cite{sewell2017, liu2023}). In contrast, SMC methods provide exact inference for SSMs and are advantageous when the number of observations in time is large. Furthermore, SMC allows both online and offline inference and estimates can be conveniently updated given additional observations.

Our proposed methodology is closely related to the surrounding literature, but differs in a few key ways. Firstly, our proposed DLSN model builds upon the SSM formulation of \cite{sarkar2006, sewell2015} but instead assumes stationarity in the latent trajectories so that properties of the network interactions are preserved through time. This is advantageous in data settings when the properties of the observed networks do not change significantly between time points. Secondly, whilst SMC methodology has previously been used in the context of latent variable network data \citep{sarkar2007, xu2014}, we present an inference procedure which does not rely on particular model forms or approximations, such as in \cite{sarkar2007, xu2014}, by appealing to methodology proposed for high-dimensional state spaces. As a consequence of this, our approach can also be easily applied to modified version of our DLSN model. Finally, we also note that our work differs from \cite{bloemreddy2018} who focus on the application for SMC to a sequence of growing graphs.

The remainder of this paper is organised as follows. In Section \ref{sec:DLSNet}, we introduce our proposed model and comment further on its context within the broader literature. Then, in Section \ref{sec:SMCandNet}, we review SMC methodology and detail our inference procedure. We present two simulation studies in Section \ref{sec:smc_sims}. The first study examines the scalability of our approach as the number of nodes or the number of observations in time grows and the second study explored the performance of our methodology under misspecification. Finally, we examine a real world data example describing classroom contacts in Section \ref{sec:realdat} and we conclude with a discussion in Section \ref{sec:smc_disc}.

\section{Dynamic Latent Space Network Modelling}
\label{sec:DLSNet}

In this section, we present a latent space model for dynamic networks in which the latent trajectories are assumed to follow a stationary first order Markov process and highlight the flexibility of this model by describing a number of possible variants. When coupled with the assumption that the latent coordinates follow a Gaussian random walk in the generative model, stationarity implies that the variance of the latent coordinate random variables does not accumulate over time. This allows network properties to be preserved over time and prevents degeneracy in which networks sampled from the model are empty given a sufficiently large time horizon.

The model outlined in Section \ref{sec:dlsn_model} is most similar to the work of \cite{sarkar2006} and \cite{sewell2015}. However, in contrast to these works, we impose stationarity on the latent trajectories. This relates to the work of \cite{he2019} who consider a latent factor model with stationarity, but differs since our formulated is based on latent Euclidean distances. A myriad of extensions have also been discussed in the literature such as the inclusion of community structures \citep{sewell2015}, weighted interactions \citep{Sewell2016105}, ranked interactions \citep{RSSC:RSSC12093}, multiple interaction types \citep{Durante2017} and bipartite networks \citep{friel2016}, and we note that our model may also be extended analogously as appropriate. For a detailed review of dynamic latent space network modelling we refer to \cite{kim2017}.

\subsection{Proposed Model and Identifiability}
\label{sec:dlsn_model}

Suppose that we observe interactions among a population of size $N$ at times $t=1,2,\dots,T$. We represent this information as a collection of time-indexed adjacency matrices $\mathcal{Y}_t = \{y_{ijt}\}_{i,j=1,2,\dots,N}$, where $y_{ijt}$ indicates the connection between nodes $i$ and $j$ at time $t$. We will consider binary connections, so that $y_{ijt} \in \{0,1\}$ where $y_{ijt}=1$ indicates the presence of the $(i,j)^{th}$ edge. Additionally, we will assume that the connections are symmetric, so that $y_{ijt}=y_{jit}$, and that there are no self ties, so that $y_{iit}=0$. 

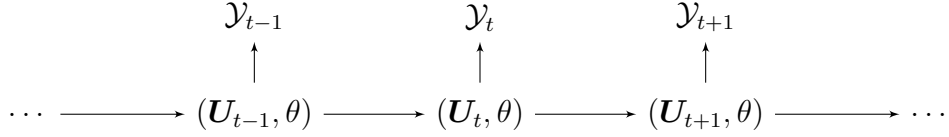
\begin{figure}[t]
  \centering
  \begin{tikzpicture}[node distance=3cm,auto,>=latex']
    \node [int, pin={[init]above:$\mathcal{Y}_{t-1}$}] (a) {($\bm{U}_{t-1}, \theta)$};
    \node [int] [left of=a, node distance=3cm] (d) {$\dots$};    
    \node [int, pin={[init]above:$\mathcal{Y}_t$}] (b) [right of=a] {($\bm{U}_t, \theta)$};
    \node [int, pin={[init]above:$\mathcal{Y}_{t+1}$}] (c) [right of=b] {($\bm{U}_{t+1}, \theta)$};
    \node [int] (end) [right of=c, node distance=3cm]{$\dots$};
    \path[->] (d) edge node {} (a);
    \path[->] (a) edge node {} (b);
    \path[->] (b) edge node {} (c);
    \draw[->] (c) edge node {} (end);
  \end{tikzpicture}
\caption{SSM for DLSNs. The observed adjacency matrices are modelled independently conditional on the latent coordinates, and the latent coordinates are modelled with a first order Markov process that is governed by static parameters $\theta$.} \label{fig:SSMNet}
\end{figure}

In the latent space framework, the probability of a connection forming is expressed as a function of $d$-dimensional coordinates associated with the nodes. We let $\bm{U}_t \in \mathbb{R}^{N \times d}$ represent the $N \times d$ matrix of latent coordinates at time $t$ where the $i^{th}$ row of $\bm{U}_t$ corresponds to the latent coordinate of the $i^{th}$ node at time $t$, $u_{it} \in \mathbb{R}^d$. Following \cite{sarkar2006, sewell2015}, we model the observations $\{ \mathcal{Y}_t \}_{t=1}^T$ using a state space model (SSM) (see Figure \ref{fig:SSMNet}), where we assume that the interactions at time $t$ occur independently conditional on the latent states $\{ \bm{U}_t \}_{t=0}^T$ which follow a first order Markov process. We further assume that the latent trajectories for each node follow independent stationary Gaussian random walks so that
\begin{align}
    p( \bm{U}_0 | \theta ) &= \prod_{i=1}^N \mathcal{N}_d\left(0, \dfrac{\sigma^2}{1-\phi^2} I_d \right)  \label{eq:u0} \\
  p( \bm{U}_t | \bm{U}_{t-1}, \theta ) &= \prod_{i = 1}^N \mathcal{N}_d( \phi {u}_{i,t-1}, \sigma^2 I_d)  &&\mbox{ for } t = 1,2, \dots, T, \label{eq:ut_transition} 
\end{align}
where $\sigma \in \mathbb{R}_{>0}$, $| \phi | < 1$ and $\theta = (\sigma, \phi)$. 

Given the latent trajectories, we can now express the probability of observing connections $\mathcal{Y}_t$. Following \cite{hoff2002}, we assume that, conditional on $\bm{U}_t$, each edge can be modelled independently via logistic regression such that nodes whose latent coordinates are close in terms of Euclidean distance are more likely to be connected. More specifically, for $t=1,2,\dots,T$, we let
\begin{align}
  p( \mathcal{Y}_t | \bm{U}_t, \alpha ) &=  \prod_{i < j} p_{ijt}^{y_{ijt}} ( 1 - p_{ijt} )^{1 - y_{ijt}} \label{eq:pygivenu} \\
  p_{ijt} &= \dfrac{1}{1 + e^{- \eta_{ijt} }} \label{eq:modellink} \\
  \eta_{ijt} &= \alpha - \| u_{it} -  u_{jt} \|, \label{eq:m_eta}
\end{align}
where $\alpha \in \mathbb{R}$ controls the global tendency for connections to form and $\| \cdot \|$ denotes the $\ell^2$-norm. 

Specifying the connection probabilities as a function of the Euclidean distance has two primary advantages. Firstly, this implies that the coordinates $\bm{U}_t$ provide an intuitive visualisation of the network at time $t$ and, secondly, this encourages transitive relationships which are often observed in social networks. Additionally, there exist many possible variants of this model. For example, we may mimic the projection model of \cite{hoff2002} by expressing \eqref{eq:m_eta} as a function of $u_i^Tu_j$ or we may adapt \eqref{eq:pygivenu} and \eqref{eq:modellink} to express non-binary connections. These adaptations will be illustrated in Section \ref{sec:realdat}.

Since \eqref{eq:pygivenu} only depends on $\bm{U}_t$ through the pairwise distances, $p(\mathcal{Y}_t | \bm{U}_t, \theta)$ is invariant to distance-preserving transformations of $\bm{U}_t$. This is a key issue associated with latent space network models and, to resolve this, many authors rely on a Procrustes transformation (for example see \cite{hoff2002}, \cite{sewell2015}) which finds the coordinates $\hat{\bm{U}}$ that minimise the sum of squared differences between $\bm{U}$ and some pre-specified reference coordinates $\bm{U}_0$. Following \cite{durante2014}, we opt to examine the model in terms of the connection probabilities which do not suffer from non-identifiability.

\section{Model Inference}
\label{sec:SMCandNet}

The model outlined in Section \ref{sec:dlsn_model} presents a natural setting for the application of SMC methodology and, in this section, we discuss the details of this connection. SMC offers a number of advantages over existing approaches. Firstly, this class of algorithms is based on recursive estimation and so SMC is more scalable than MCMC methods when the number of time steps $T$ is large. Secondly, in contrast to variational Bayes techniques, SMC does not require model approximations. Finally, SMC methods facilitate both online and offline inference and allow posterior samples to be updated given additional observations. We provide a brief overview of SMC in Section \ref{sec:pf_background} and present our estimation procedure in Section \ref{sec:dlsn_est}. Since the performance of standard SMC methods degrades as the dimension of the state space increases, we focus our attention on high-dimensional SMC techniques that are more appropriate for the large network setting.

\subsection{Background: SMC review}
\label{sec:pf_background}

SMC methods are a class of simulation-based algorithms designed to iteratively estimate a posterior distribution (see \cite{doucet2009}, \cite{Lopes2011} and \cite{Doucet2001}). This section provides a review of SMC methodology for a generic SSM and can be skipped by the familiar reader. We split our discussion into state estimation in Section \ref{sec:pf_ssm}, where we focus on particle filtering, and static parameter estimation in Section \ref{sec:pf_thetaest}. To motivate the algorithm considered in Section \ref{sec:dlsn_state_est}, we briefly discuss high-dimensional SMC methodology in Section \ref{sec:smc_hd}.

\subsubsection{General SSM}
\label{sec:general_ssm}

In this section we review methodology for a general SSM governed by
\begin{align}
  &X_0 \sim \mu_{\theta}(x_0), \nonumber \\  
  &X_t | X_{t-1} = x_{t-1} \sim f_{\theta}(x_t | x_{t-1} ), &&\mbox{ for } t=1,2,\dots,T,  \label{eq:SSM} \\
  &Z_t | X_t = x_t \sim g_{\theta}(z_t | x_t), &&\mbox{ for } t=1,2,\dots,T, \nonumber   
\end{align}
where $\theta$ is a vector of parameters. Note that the model in Section \ref{sec:dlsn_model} is a special case of \eqref{eq:SSM} with $\theta = (\alpha, \phi, \sigma)$, $Z_t = \mathcal{Y}_t$ and $X_t = \bm{U}_t$ and densities as specified in \eqref{eq:u0}, \eqref{eq:ut_transition} and \eqref{eq:pygivenu}.

\subsubsection{Generic Particle Filter}
\label{sec:pf_ssm}

A particle filtering (PF) scheme estimates the latent states $\{x_t\}_{t=0}^T$ given observations $\{ z_t \}_{t=1}^T$ by sequentially targeting the filtering densities $\{ p_{\theta}(x_t|$ $z_{1:t}) \}_{t = 1}^T$. In general, we cannot obtain analytic expressions for the filtering densities and so a PF scheme instead relies on approximations obtained via importance sampling (IS). However we note that, in the special case when $f_{\theta}(x_{t} | x_{t-1})$ and $g_{\theta}(z_{t} | x_t)$ are Gaussian, we may determine $p_{\theta}(x_t|$ $z_{1:t})$ analytically and this scheme is the well-known Kalman filter \citep{Kalman1960}.

A generic PF scheme for approximating $\{ p_{\theta}(x_t | z_{1:t}) \}_{t=1}^T$ is outlined in Algorithm \ref{alg:PF}, where we let $\{x_t^{(i)}, w_t^{(i)} \}_{t=1}^M$ denote the particle approximation to $p_{\theta} (x_t | z_{1:t})$. In this notation $x_t^{(i)}$ and $w_t^{(i)}$ represent the $i^{th}$ particle and $i^{th}$ weight, respectively. This scheme updates the particles at time $t-1$ according to a proposal distribution $q_{\theta}(x_t | x_{t-1}, z_t)$ and then adjusts the weights to account for the discrepancy between the $t^{th}$ and $(t-1)^{th}$ filtering densities. Algorithm \ref{alg:PF} also contains a resampling step, denoted as $\mathcal{F}( \cdot | \cdot)$, to mitigate particle degeneracy in which the weights concentrate onto a small number of particles causing the quality of the approximations to degrade. We refer to \cite{Douc2005} and \cite{doucet2009} for details of standard resampling schemes. 

\begin{algorithm}[t]
  \caption{Generic Particle Filter to obtain estimates $\{x_t^{(i)}, w_t^{(i)} \}_{i=1}^M$ of $p_{\theta} (x_t | z_{1:t})$, for $t=1,2,\dots,T$.} \label{alg:PF}
  \begin{algorithmic}
    \STATE $\bullet$ \textit{Iteration $t=0$:}
    \STATE  Sample $M$ particles $\{ x_0^{(i)}  \}_{i=1}^M$ from $\mu_{\theta}(\cdot)$ and assign weights $w_0^{(i)} = 1/M$.
    \STATE $\bullet$ \textit{Iteration $t=1, \dots, T$:}
    \STATE Assume particles $\{ x_{t-1}^{(i)}  \}_{i=1}^M$ with weights $\{ w_{t-1}^{(i)}  \}_{i=1}^M$ that approximate $p_{\theta}(x_{t-1} | z_{1:t-1})$. 

    \STATE a) Sample parent indices $\{a_{t-1}^{(i)}\}_{i=1}^M$ from $\mathcal{F}( \cdot | W_{t-1})$, where $W_{t-1} = (w_{t-1}^{(1)},w_{t-1}^{(2)}, \dots, w_{t-1}^{(M)} )$. 
   
    \STATE b) Propagate the particles according to $x_t^{(i)} \sim q_{\theta}\left(\cdot |z_t, x_{t-1}^{a_{t-1}^{(i)}}\right) $, for $i=1,2,\dots,M$, and set $x_{1:t}^{(i)} = \left(x_{1:t-1}^{a_{t-1}^{(i)}}, x_t^{(i)} \right)$
   
    \STATE c) Calculate the weights
    \begin{align*}
    w_t^{(i)} \propto {p_{\theta}\left(x_{1:t}^{(i)}, z_{1:t}\right)} \bigg/ {q_{\theta} \left(x_t^{(i)}|z_t, x_{1:t-1}^{a_{t-1}^{(i)}}\right) p_{\theta}\left(x_{1:t-1}^{a_{t-1}^{(i)}}, z_{1:t-1}\right) }
  \end{align*}
  for $i=1,2,\dots,M$ and normalise.
  \end{algorithmic}
\end{algorithm}

Many well-established particle filtering algorithms can be obtained as a special case of the scheme in Algorithm \ref{alg:PF}. For instance, we may obtain the standard SIR filter of \cite{Gordon1993} by taking $q_{\theta}(\cdot | x_{t-1}, z_t) = f_{\theta}( \cdot | x_{t-1})$ and $w_t = g_{\theta}(z_t | \cdot)$. This filter has the advantage of being simple to implement, however the particles are propagated according to the model dynamics and so do not incorporate information about future observations. It has been shown that the proposal which obtains optimal performance in terms of the variance of the importance weights is \citep{doucet2000} $q_{\theta}(x_t | z_t, x_{t-1}) = p_{\theta}(x_t | z_t, x_{t-1} )$. However, for many practical applications, we cannot find a tractable expression for this proposal. Alternative approaches, such as the auxiliary particle filter (APF) of \cite{Pitt1999}, instead aim to approximate $p(x_t | z_t, x_{t-1})$. 

To assess the performance of a PF, we can approximate the the effective sample size (ESS) at time $t$ as $\widehat{ESS} = {1} \big/ { \sum_{i=1}^M (w_t^{(i)} )^2 }$. Additionally, a PF allows us to obtain an estimate of the marginal likelihood $p_{\theta}(y_{1:T})$ from the importance weights using
\begin{align}
  \hat{p}_{\theta}(y_{1:T}) = \prod_{t=1}^T \left[ \dfrac{1}{M} \sum_{i=1}^M w_t^{(i)} \right]. \label{eq:py_pf}
\end{align}
Evaluating \eqref{eq:py_pf} does not incur an additional computational cost and this estimator has been shown to be unbiased (see Theorem 7.4.2 in \cite{Moral2004}).


\subsubsection{Static Parameter Estimation}
\label{sec:pf_thetaest}

The procedure in Algorithm \ref{alg:PF} estimates the latent states $\{x_t\}_{t=1}^T$ conditional on known static parameters $\theta$. However, in practice, we also wish to estimate static parameters $\theta$, given by $(\sigma, \phi, \alpha)$ in our application. The existing literature can be divided into offline and online approaches, and we will briefly discuss these below. For a more in depth discussion, we refer the reader to \cite{Kantas2009}, \cite{gao2012}, \cite{Kantas2015} and \cite{Lopes2011}.

Offline estimation can be considered within the Bayesian or Frequentist paradigm. In the Bayesian setting, particle MCMC \citep{Andrieu2010} methods allow joint state and parameter estimation by utilising an SMC approximation of the the likelihood which leaves the target distribution invariant. This avoids calculation of complex proposal distributions and, due to the crucial role of the estimation of the marginal likelihood, attention has been focused on developing PF schemes which improve the quality of this estimate (for example, see \cite{ala-luhtala2016}). Alternatively, \cite{Chopin2013} develop an iterated procedure which relies on nested particle filters to estimate the state and parameters jointly in the Bayesian setting. From the Frequentist perspective, estimating $\theta$ is viewed as a likelihood maximisation problem (see \cite{malik2011} and \cite{hurzeler2001}). This has been addressed via gradient ascent in which estimates of $\theta$ are updated according to the gradient of the log-likelihood (for example, see \cite{ionides2011}). \cite{poyiadjis2011} consider estimating the score and information matrix, which can be used within gradient ascent, and \cite{nemeth2016} extend this to develop a procedure which has a linear computational cost. Alternatively, $\theta$ can be estimated using an expectation maximisation procedure as in \citep{dempster1977}. 

In the online setting, a natural approach for estimating $\theta$ would be to find a particle approximation to the joint density $p(x_t, \theta | z_{1:t})$, similar to the procedure for estimating $\{ x_t\}_{t=1}^T$ outlined in Algorithm \ref{alg:PF}. However, since $\theta$ does not evolve in time, the particle set will degenerate under repeated resampling. Online estimation of $\theta$ therefore presents a challenging task and remains an open problem in the literature. Several Bayesian strategies have been proposed, such as \cite{Gordon1993}, who include artificial dynamics to reduce the degeneracy, and \cite{gilks2001}, who rely on MCMC kernels to add diversity to the particle set. Other approaches include practical filtering (see \cite{polson2008}), kernel approximations (see \cite{Liu2001}) and estimating $\theta$ using sufficient statistics (see \cite{Storvik2002}, \cite{Fearnhead2002} and \cite{Carvalho2010}). More recently, \cite{crisan2018} introduced a procedure which relies on nested particle filters, similarly to the approach of \cite{Chopin2013}. In the Frequentist setting, $\theta$ can be estimated via likelihood maximisation. For example, see \cite{legland1997}, \cite{poyiadjis2011} and \cite{nemeth2016}, for methodology which relies on gradient ascent. Alternatively, \cite{Cappe2011} consider estimation via an expectation maximisation procedure.

\subsubsection{High-Dimensional Particle Filtering}
\label{sec:smc_hd}

It is well-understood that the performance of SMC methods degrade as the dimension of the state space increases \citep{bengtsson2008, snyder2008, beskos2014a, beskos2014b} and it has been shown that the number of particles must increase exponentially with the state dimension to avoid particle degeneracy \citep{snyder2015}. Developing methodology for high-dimensional state spaces remains an open problem in the literature, and in recent years a body of work has developed to address this issue.

A popular strategy involves the utilisation of MCMC techniques, such as in \cite{Godsill2001} where MCMC moves are introduced to mitigate against particle degeneracy. More broadly, MCMC methodology has appeared as part of sequential MCMC schemes \citep{Khan2005, Septier2009, Brockwell2012} and this idea has been explored further in, for example, \cite{nguyen2015} and \cite{septier2016}. Other authors have considered approximate algorithms, such as the ensemble Kalman filter (see \cite{katzfuss2016}) which approximates a Kalman filter through a sample, or `ensemble', from the target distribution. Alternatively, \cite{rebeschini2013} explore local particle filters such as the block particle filter in which the state space is partitioned into non-overlapping blocks. Finally, \cite{akyildiz2019} generalise the nudging approach, where a subset of particles are artificially moved towards the observation (see also \cite{leeuwen2010} and \cite{ades2015}).

Several algorithms have also been developed for particular classes of models. \cite{beskos2014} introduce the space-time filter which assumes that the likelihood can be decomposed so that knowledge of the observations can be incrementally incorporated. \cite{wigren2018} introduce artificial process noise to improve the performance of the particle filter in high-dimensions for a class of state space models. \cite{naesseth2016} consider a nested approach in which an additional particle filter is introduced to approximate the proposal distribution relied upon in the fully adapted filter of \cite{Pitt1999}. \cite{park2019} introduce artificial intermediary states in order to handle high-dimensional state spaces. We note that in our application of interest the form of our network model restricts which of these algorithms are appropriate. Finally, for a selective review of high-dimensional particle filtering algorithms, we refer the reader to \cite{septier2015}. 

\subsection{Estimation Procedure}
\label{sec:dlsn_est}

In this section we detail our inference procedure for the DLSN model presented in Section \ref{sec:dlsn_model}. In our setting, the dimension of the state space is given by the product over the number of nodes $N$ and the latent dimension $d$, namely $N \times d$. Hence, even for moderate $N$, we must consider methodology for high-dimensional state spaces. We discuss our procedure for state and parameter estimation in Sections \ref{sec:dlsn_state_est} and \ref{sec:dlsn_prm_est}, respectively.

\subsubsection{State Estimation}
\label{sec:dlsn_state_est}

\begin{figure}[h!]
  \centering
  \begin{tikzpicture}[node distance=1.75cm,auto,>=latex']
    \node [int, pin={[init]above:$\mathcal{Y}_t$}] (a) {$\bm{U}_{\tau_{t,0}}$};
    \node [int] [left of=a, node distance=1.75cm] (d) {$\dots$};    
    \node [int] (b) [right of=a] {$\bm{U}_{\tau_{t,1}}$};
    \node [int] (c) [right of=b] {$\bm{U}_{\tau_{t,2}}$};
    \node [int] (e) [right of=c] {$\dots$};
    \node [int] (f) [right of=e] {$\bm{U}_{\tau_{t,S-1}}$};
    \node [int] (g) [right of=f, pin={[init]above:$\mathcal{Y}_{t+1}$}] {$\bm{U}_{\tau_{t,S}}$}; 
    \node [int] (end) [right of=g, node distance=1.75cm]{$\dots$};
    \path[->] (d) edge node {} (a);
    \path[->] (a) edge node {} (b);
    \path[->] (b) edge node {} (c);
    \path[->] (c) edge node {} (e);
    \path[->] (e) edge node {} (f);
    \path[->] (f) edge node {} (g);
    \draw[->] (g) edge node {} (end);
  \end{tikzpicture}
  \caption{Intermediary steps $\{\bm{U}_{\tau_{t,s}}\}_{s=0}^S$ between observations $\mathcal{Y}_t$ and $\mathcal{Y}_{t+1}$, where $\bm{U}_{\tau_{t,0}} = \bm{U}_t$ and $\bm{U}_{\tau_{t,S}} = \bm{U}_{t+1}$.} \label{fig:girfsteps}
\end{figure}
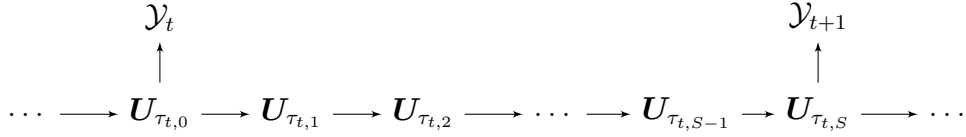

To estimate the latent coordinates we rely on the Guided Intermediate Resampling Filter (GIRF) of \cite{park2019}. This scheme introduces artificial intermediary states in order to guide particles to regions of the state space with high probability via a guide function, typically taken as the predictive likelihood. We opt to use this scheme since it is appropriate for our modelling assumptions, is straightforward to implement and can be adapted to model variants, such as those described in Section \ref{sec:DLSNet}. It is important to stress that the dependence in our likelihood \eqref{eq:pygivenu} makes it challenging to consider many alternative high-dimensional particle filtering schemes proposed in the literature, such as \citep{rebeschini2013, beskos2014}.

Our goal is to estimate $\{ \bm{U}_t\}_{t=1}^T$ given the observed connections $\{ \mathcal{Y}_t\}_{t=1}^T$. For each pair of observations $\{\mathcal{Y}_{t}, \mathcal{Y}_{t+1}\}_{t=1}^{T-1}$, the GIRF introduces $S-1$ intermediary time steps which we denote by $\{\tau_{t,s} \}_{s=0}^S$. Then, we rely upon the following decomposition for the latent transitions.
\begin{align}
  p_{\theta}(\bm{U}_{t+1} | \bm{U}_{t}) = \tilde{p}_{\theta}(\bm{U}_{\tau_{t,1}} | \bm{U}_{t}) \tilde{p}_{\theta}(\bm{U}_{\tau_{t,2}} | \bm{U}_{\tau_{t,1}}) \dots \tilde{p}_{\theta}(\bm{U}_{t+1} | \bm{U}_{\tau_{t,S-1}}), \label{eq:kerneldiv} 
\end{align}
where $\{ \tau_{t,s} \}_{s=0}^S$ satisfy $\tau_{t,0} := t < \tau_{t,1} < \dots < \tau_{t,S-1} < \tau_{t,S} := t+1$, and $\tau_{t,0}$ and $\tau_{t,S}$ correspond to the times $t$ and $t+1$, respectively. Figure \ref{fig:girfsteps} provides a depiction of the intermediary states and a derivation of $\tilde{p}_{\theta}(\bm{U}_{s} | \bm{U}_{\tau_{t,s-1}})$ for the transition equation in \eqref{eq:ut_transition} is given in \ref{app:trans_eq_ptilde}. Note that, for this intermediary transition to be valid, we require $\phi \in (0,1)$ but this does not affect the expressivity of our model since connections are modelled in terms of the latent distances.

At each intermediary time step, the particles are weighted according to an assessment function $\nu_{\tau_{t,s}}(u)$ which guides particles towards future observations. This function must be specified so that $\nu_{\tau_{0,0}}(u)=1$ and $\nu_{\tau_{T,0}}(u)=p(\mathcal{Y}_T|u)$, and we will discuss particular choices below. Given this function, particles at time step $\tau_{t,s}$ are then weighted according to
\begin{align}
  \omega_{\tau_{t,s}}\left( \bm{U}_{\tau_{t,s}}, \tilde{\bm{U}}_{\tau_{t,s-1}} \right) =
  \begin{cases} 
      \dfrac{ \nu_{\tau_{t,s}}( \bm{U}_{\tau_{t,s}}) }{\nu_{\tau_{t,s-1}}( \tilde{\bm{U}}_{\tau_{t,s-1}})} & \mbox{ if } \tau_{t,s-1} \not\in 1:T\\
      \dfrac{ \nu_{\tau_{t,s}}( \bm{U}_{\tau_{t,s}}) }{\nu_{\tau_{t,s-1}}( \tilde{\bm{U}}_{\tau_{t,s-1}})} p( \mathcal{Y}_t | \tilde{\bm{U}}_{\tau_{t,s-1}}) & \mbox{ if } \tau_{t,s-1} \in 1:T 
   \end{cases}. \label{eq:weights}
\end{align}

Using \eqref{eq:weights}, we see that the likelihood $\ell (\mathcal{Y}_{1:T}) = \mathbb{E} \left[ \prod_{t=1}^T p(\mathcal{Y}_t | \bm{U}_t)\right]$ can be approximated by $\hat{\ell} = \prod_{t=0}^T \prod_{s=1}^S \dfrac{1}{M} \sum_{i=1}^M \omega_{\tau_{t,s}}\left( \bm{U}_{\tau_{t,s}}^{(i)}, \tilde{\bm{U}}_{\tau_{t,s-1}}^{(i)} \right)$. 

\begin{algorithm}[t]
  \caption{GIRF for DLSN} \label{alg:girf}
  \begin{algorithmic}{
    \STATE \textbf{Initialise}: $L = 0$, sample $\tilde{\bm{U}}_{\tau_{0,0}}^{(i)} \sim p(\cdot| \theta )$ for $i \in 1:M$
    \STATE  \textbf{For} $t = 0:T-1$
    \STATE \hspace{1cm} \textbf{For} $s \in 1:S$
    \STATE \hspace{1cm} $\bm{U}_{\tau_{t,s}}^{(i)} \sim p( \cdot | \tilde{\bm{U}}_{\tau_{t,s-1}}^{(i)}, \theta)$ for $i \in 1:M$ \\
    \STATE \hspace{1cm} $w_{\tau_{t,s}}^{(i)} = \omega_{\tau_{t,s}}(\bm{U}_{\tau_{t,s}}^{(i)}, \tilde{\bm{U}}_{\tau_{t,s-1}}^{(i)})$ for $i \in 1:M$ \\
    \STATE \hspace{1cm} $L = L + \log( \sum_j w_{\tau_{t,s}}^{(i)})$ \\
    \STATE \hspace{1cm} Sample $b^{(i)}$ such that $\mathbb{P}(b^{(i)}=j) \propto w^{(j)}_{\tau_{t,s}}$ for $i \in 1:M$ \\
    \STATE \hspace{1cm} Set $\tilde{\bm{U}}_{\tau_{t,s}}^{(i)} = \bm{U}_{\tau_{t,s}}^{(b^i)}$
    \STATE \hspace{1cm} \textbf{End} \\
    \STATE \textbf{End} 
  }
  \end{algorithmic}
\end{algorithm}

Algorithm \ref{alg:girf} details the GIRF for the DLSN model in Section \ref{sec:dlsn_model} where $L = \log \hat{\ell}$. To implement this procedure we must specify $\nu_{\tau_{t,s}}(x)$ and the number of intermediary states $S$. \cite{park2019} suggest choosing $\nu_{\tau_{t,s}}(u)$ so that particles are guided towards $B$ future observations and then demonstrate that $S$ should scale linearly according to the dimension of the latent states. In our setting, a straightforward option is to take $\nu_{\tau_{t,s}}(\bm{U}) = p(\mathcal{Y}_{t+1} | \bm{U})$ or, to incorporate $B$ future observations,  
\begin{align}
  \nu_{\tau_{t,s}}(\bm{U}) = \prod_{b=1}^{\min\{B, T-t \}} \left( \nu_{\tau_{t,s}, \tau_{t+b}}(\bm{U}) \right)^{\eta_{\tau_{t,s}, \tau_{t+b}}}, \label{eq:uwithB}
\end{align}
where $\nu_{\tau_{t,s}, \tau_{t+b}}(\bm{U})$ approximates $p_{\mathcal{Y}_{t+b}|\bm{U}_{\tau_{t,s}}}(\mathcal{Y}_{t+b} | \bm{U})$ and $\eta_{\tau_{t,s}, \tau_{t+b}}$ controls the contribution of $\nu_{\tau_{t,s}, \tau_{t+b}}(\bm{U})$ in the assessment function. We let
\begin{align}
\eta_{\tau_{t,s}, \tau_{t+b}} = 1 - \dfrac{(bS - s)}{S \left[ (t+b) - \max(t+b-B,0) \right]}  \label{eq:eta_uwithB}
\end{align}
so that the contribution of observations decreases as a function of distance from $\tau_{t,s}$. This ensures that the potentially less accurate approximations have a smaller contribution to $\nu_{\tau_{t,s}}(\bm{U})$. It may be possible to approximate $p_{\mathcal{Y}_{t+b}|\bm{U}_{\tau_{t,s}}}(\mathcal{Y}_{t+b} | \bm{U})$ via simulation, however this will be computationally expensive for our setting. Instead we take $\nu_{\tau_{t,s}, \tau_{t+b}}(\bm{U}) = g(\mathcal{Y}_{t+b}| m_{\tau_{t,s}}(\bm{U}))$, where $m_{\tau_{t,s}}(\bm{U}) = \mathbb{E}\left[ \bm{\bm{U}}_{t+b} | \bm{\bm{U}}_{\tau_{t,s}} = \bm{U} \right] = \phi \bm{U}$, which can be conveniently calculated. 

\ref{app:check_girf} provides an example of the GIRF on simulated data with known $\theta$. From this, we find that the choice of $S$ significantly impacts the performance of the filter, with larger values improving the estimated ESS. We also found little difference in practice between choosing $p(\mathcal{Y}_{t+1} | \bm{U})$ versus \eqref{eq:uwithB} as the assessment function for this example. 

\subsubsection{Parameter Estimation}
\label{sec:dlsn_prm_est}

We now consider estimation of static parameters $\theta = (\sigma, \phi, \alpha)$ using the gradient ascent approach of \cite{nemeth2016}. This allows both online and offline estimation, is flexible to variants of the model and has been shown to exhibit favourable performance when compared with other well known algorithms. We begin by discussing their approach for offline estimation of $\theta$.  

A gradient ascent procedure obtains the $k^{th}$ estimate of $\theta$ by taking
\begin{align}
  \theta_k = \theta_{k-1} + \gamma_k \nabla \log p \left( \mathcal{Y}_{1:T} | \theta \right) |_{\theta = \theta_{k-1}}, \label{eq:thetagradupdate}
\end{align}
where $\gamma_k$ is a sequence of decreasing steps such that $\sum_k \gamma_k = \infty$ and $\sum_k \gamma_k^2 < \infty$. A typical choice for this sequence is $\gamma_k = k^{-\alpha}$ with $0.5 < \alpha < 1$.

Following \cite{nemeth2016}, we can rely on a particle approximation of the score $s_T = \nabla \log p( \mathcal{Y}_{1:T} | \theta)   |_{\theta = \theta_{k-1}}$ to evaluate \eqref{eq:thetagradupdate}. Details of this approximation are given in Algorithm \ref{alg:gradprm}, where $s_t = \nabla \log p( \mathcal{Y}_{1:t} | \theta)$. An implementation of Algorithm \ref{alg:gradprm} returns an estimate of $s_T$ which can then be used to update $\theta$ according to \eqref{eq:thetagradupdate}. An offline procedure therefore requires $N_{\theta}$ runs of the PF, and we implement this by using the GIRF as our PF in Algorithm \ref{alg:gradprm}.

An online implementation can be obtained by assuming
\begin{align}
  \nabla \log p(\mathcal{Y}_t | \mathcal{Y}_{1:t-1}, \theta_t) \approx  \nabla \log p(\mathcal{Y}_{1:t} | \theta_t ) - \nabla \log p(\mathcal{Y}_{1:t-1} | \theta_{t-1} ), \label{eq:scoreapprox}
\end{align}
which typically holds for small changes between $\theta_{t-1}$ and $\theta_t$. Then, at the $t^{th}$ iteration, we approximate the score $s_T$ by $\nabla \log \hat{p} (\mathcal{Y}_t |$ $\mathcal{Y}_{1:t-1}, \theta_t ) = s_t - s_{t-1}$, where $s_t$ is obtained via Algorithm \ref{alg:gradprm}. The details of this are provided in Algorithm \ref{alg:online_theta_est}.

\begin{algorithm}[t]
  \caption{Rao-Blackwellised Score} \label{alg:gradprm}
  \begin{algorithmic}
    \STATE \textbf{Initialise}: set $m_0^{(i)}=0$ for $i=1,2,\dots,M$, $s_0=0$.
    \STATE \textbf{For} $t=1,2,\dots,T$
    \STATE \hspace{.75cm} 1) Run one iteration of a PF to obtain $\{\bm{U}_t^{(i)}\}_{i=1}^M, \{a_{t-1}^{(i)}\}_{i=1}^M$ and 
    \STATE \hspace{1.25cm} $\{ w_t^{(i)} \}_{i=1}^M$ (see Algorithm \ref{alg:PF})
    \STATE \hspace{.75cm} 2) Update the mean approximation \\
    \hspace{1.25cm} $m_t^{(i)} = \lambda m_{t-1}^{(a_{t-1}^{(i)})} + (1-\lambda) S_{t-1} + \nabla \log p(\mathcal{Y}_t | \bm{U}_t^{(i)}, \alpha) $ \\
    \STATE \hspace{8cm} $+ \nabla \log p(\bm{U}_t^{(i)} | \bm{U}_{t-1}^{(a_{t-1}^{(i)})}, \theta )  $ \\
    \hspace{.75cm} 3) Update the score vector  $s_t = \sum_{i=1}^M w_t^{(i)} m_t^{(i)}$ 
    \STATE \textbf{End}
  \end{algorithmic}
\end{algorithm}

\begin{algorithm}[t]
  \caption{Online parameter estimation within the GIRF} \label{alg:online_theta_est}
  \begin{algorithmic}
    \STATE \textbf{Initialise}: set $s_0=0, L=0, \theta_0, m_0^{(i)}=0$ and $\tilde{\bm{U}}_{\tau_{0,0}}^{(i)} \sim p(\cdot| \theta)$ for $i=1,2,\dots,M$. 
    \STATE \textbf{For} $t=1,2,\dots,T$
    \STATE \hspace{.75cm} 1) Run $S$ intermediary steps of the GIRF to obtain $\{\bm{U}_t^{(i)}\}_{i=1}^M, $ \\
    \STATE \hspace{1.25cm} $\{a_{t-1}^{(i)}\}_{i=1}^M$ and $\{ w_t^{(i)} \}_{i=1}^M$ (innermost loop in Algorithm \ref{alg:girf}) \\
    \STATE \hspace{.75cm} 2) Update the mean approximation \\
    \hspace{1.25cm} $m_t^{(i)} = \lambda m_{t-1}^{(a_{t-1}^{(i)})} + (1-\lambda) S_{t-1} + \nabla \log p_{\theta_{t-1}}(\mathcal{Y}_t | \bm{U}_t^{(i)}, \alpha) $ \\
    \hspace{8cm} $+ \nabla \log f_{\theta_{t-1}}(\bm{U}_t^{(i)} | \bm{U}_{t-1}^{(a_{t-1}^{(i)})}, \theta ) $ \\
    \hspace{.75cm} 3) Update the score vector $s_t = \sum_{i=1}^M w_t^{(i)} m_t^{(i)}$ \\ 
    \hspace{.75cm} 4) Update theta according to $\theta_t = \theta_{t-1} + \gamma_k(s_t - s_{t-1})$\\
    \STATE \textbf{End}
  \end{algorithmic}
\end{algorithm}

Since we require $\sigma > 0$ and $ \phi \in (0,1)$ we opt to estimate $\tilde{\theta} = (\tilde{\sigma}, \tilde{\phi}, \alpha)$ where
\begin{align}
  \tilde{\sigma} = \log(\sigma) \in \mathbb{R} \hspace{1cm} \mbox{ and } \hspace{1cm} \tilde{\phi} = \log \left( \dfrac{\phi}{1-\phi} \right) \in \mathbb{R}. \label{eq:tilde_sigandphi}
\end{align}
Details of the necessary gradient calculations and parameter initialisations are provided in \ref{app:theta_gradients_alt} and \ref{app:theta_init}, respectively.

\section{Performance for Simulated Examples}
\label{sec:smc_sims}

We now explore the performance of the GIRF when additionally estimating static parameters, as detailed in Section \ref{sec:dlsn_est}. First, we will consider the performance of this approach when the data are simulated according to different mechanisms in Section \ref{sec:sims_altscen}. Then, we will consider the scalability of this approach as $N$ and $T$ increase in Section \ref{sec:sims_scalability}.

\subsection{Simulated Data}
\label{sec:sims_altscen}

\begin{figure}[t!]
  \begin{subfigure}[t]{\textwidth}
    \centering
  \includegraphics[trim={0 6.5cm 0 1cm}, clip, width=\textwidth]{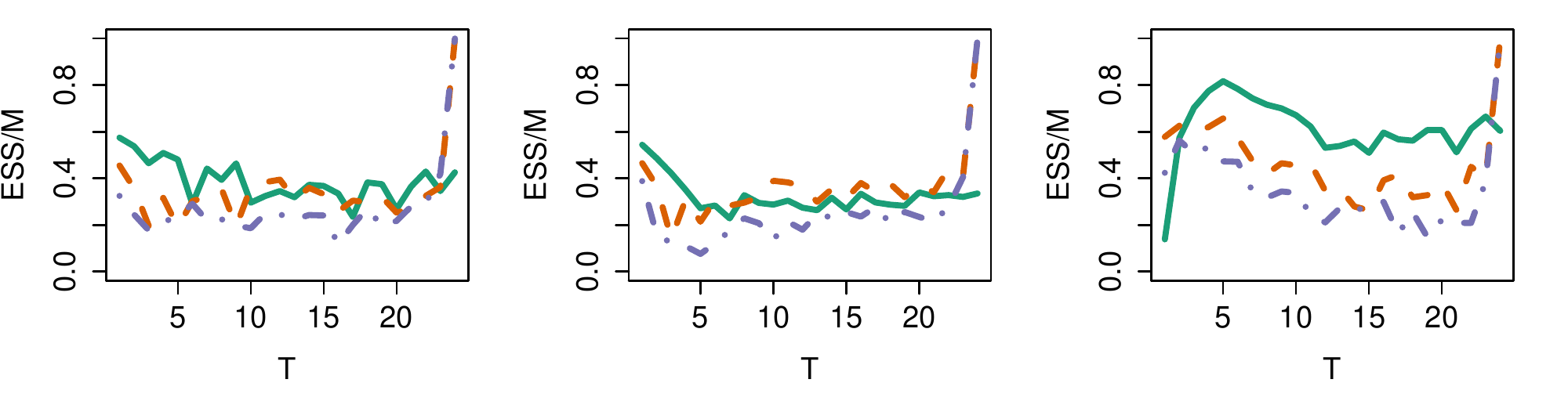}
  \caption{Effective sample size.} \label{fig:cases_online_ess}
  \end{subfigure}
  \begin{subfigure}[t]{\textwidth}
    \centering
  \includegraphics[trim={0 0cm 0 7cm}, clip, width=\textwidth]{cases_online_ess_mse}
  \caption{Mean square error in probability.} \label{fig:cases_online_mse}
  \end{subfigure}
  \caption{Figure \ref{fig:cases_online_ess} shows the effective sample size and Figure \ref{fig:cases_online_mse} shows the mean square error in probability (see \eqref{eq:mse_prob}) for the online estimation procedure run with $B=1$ (green, solid), $B=2$ (orange, dashed) and $B=3$ (purple, dot-dashed) look ahead steps. For each figure, left, middle and right correspond to data simulated according to models \ref{item:type1}, \ref{item:type2} and \ref{item:type3}, respectively.} \label{fig:online_cases_ess_mse}
\end{figure}

\begin{figure}[t!]
  \begin{subfigure}[t]{\textwidth}
    \centering
  \includegraphics[trim={0 6.5cm 0 1cm}, clip, width=\textwidth]{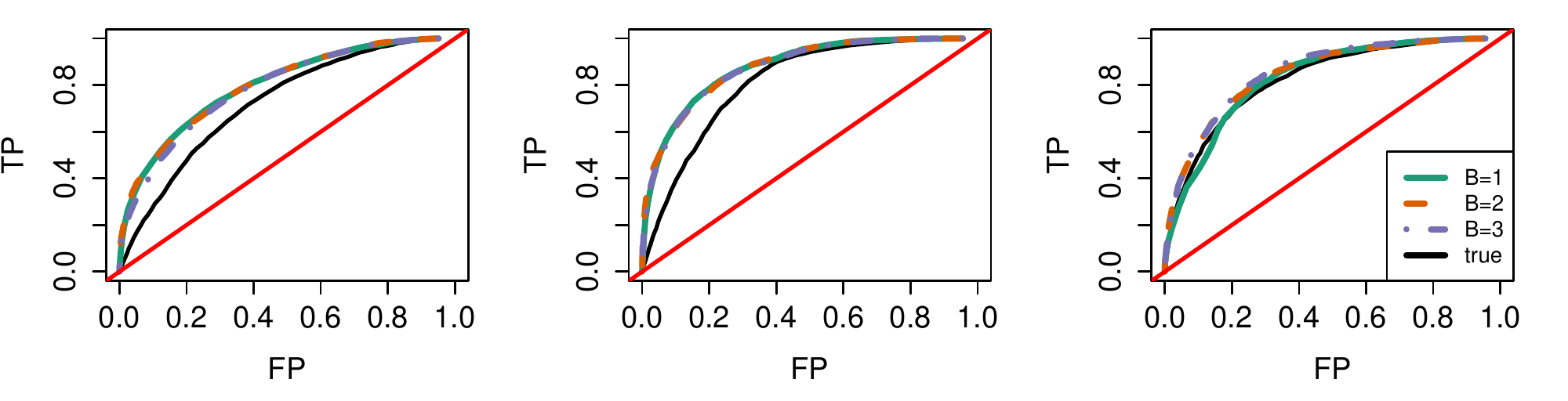}
  \caption{ROC curve for observations at time 1 to time $T-1$.} \label{fig:cases_online_rocall}
  \end{subfigure}
  \begin{subfigure}[t]{\textwidth}
    \centering
  \includegraphics[trim={0 0cm 0 7cm}, clip, width=\textwidth]{cases_online_roc}
  \caption{ROC curve for observations at time $T$.} \label{fig:cases_online_rocpred}
  \end{subfigure} 
  \caption{Figures \ref{fig:cases_online_rocall} and \ref{fig:cases_online_rocpred} show the ROC curves for observations 1 to $T-1$ and for the predicted probabilities at time $T$, respectively. Each figure reports the ROC for the online estimation procedure run with $B=1$ (green, solid), $B=2$ (orange, dashed) and $B=3$ (purple, dot-dashed) look ahead steps. For each figure, left, middle and right correspond to data simulated according to models \ref{item:type1}, \ref{item:type2} and \ref{item:type3}, respectively. The line $y=x$ is shown in red and the ROC curve for the true probabilities is shown in black.} \label{fig:online_cases_rocs}
\end{figure}

\begin{figure}[t!]
  \centering
  \includegraphics[trim={0 0cm 0 1.cm}, clip, width=\textwidth]{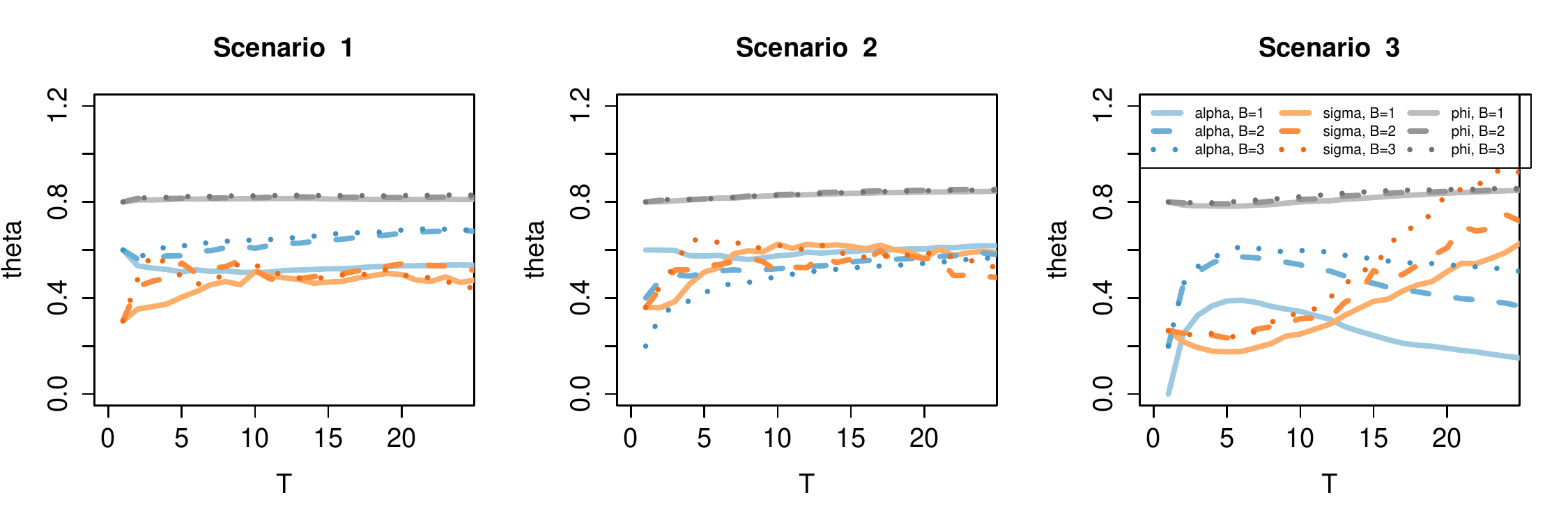}
  \caption{Estimates of $\alpha$ (blue), $\sigma$ (orange) and $\phi$ (grey) obtained via online procedure with $B=1$ (solid), $B=2$ (dashed), $B=3$ (dot-dashed). Left, middle and right figures correspond to the data generated according to \ref{item:type1}, \ref{item:type2} and \ref{item:type3}, respectively.} \label{fig:theta_online_cases}
\end{figure}

In this section our focus will be on the performance of the GIRF for networks simulated according to the following data generating mechanisms. Throughout, we fix $N=30$, $d=2$ and $T=25$.
\begin{enumerate}
  \myitem[(S1)] \label{item:type1} Simulate from the model in Section \ref{sec:dlsn_model} with $\alpha=0.75$, $\sigma=0.4$ and $\phi=0.9$.
  \myitem[(S2)] \label{item:type2} Partition the nodes into two groups with latent centres $\mu_c \in \mathbb{R}^d$, for $c \in \{1,2\}$. Suppose that the $i^{th}$ node belongs to group $c$, then we let
  \begin{align}
    u_{it}^c = (1-q)u_{i,t-1}^c + q \mu_c + \epsilon
  \end{align}
where $q \in (0,1)$ and $\epsilon \sim \mathcal{N}_d(0, \sigma^2 I_d)$. We take $\alpha=0.75, \sigma=0.4, q=0.25, \mu_1 = (2,0)$ and $\mu_2= (-2,0)$, and note that this is similar to the model of \cite{sewell2017}.
  \myitem[(S3)] \label{item:type3} Let the density $\alpha$ decrease over time, so that \eqref{eq:m_eta} is replaced by
  \begin{align}
    \eta_{ijt} = \alpha_t - \| u_{it} -  u_{jt} \|
  \end{align}
where $\bm{\alpha} = (\alpha_1, \alpha_2, \dots, \alpha_T)$ is a decreasing sequence sequence starting at $2$ and ending at $-2$. We also take $\sigma=0.4$ and $\phi=0.9$.
\end{enumerate}
\ref{item:type1} corresponds to simulating from the generative model, \ref{item:type2} corresponds to simulating from a variant of the generative model in which the nodes belong to one of two communities that evolve jointly, and \ref{item:type3} corresponds to simulating from a variance of the generative model in which the density of the networks evolves from more to less connected. Note that \ref{item:type1} allows us to verify our approach is behaving as expected, whereas \ref{item:type2} and \ref{item:type3} allow us to explore the performance of our approach under two specific forms of model misspecification.

For each of \ref{item:type1}, \ref{item:type2} and \ref{item:type3}, we fit our model using both an online and offline procedure. We take $S=1.5N, M=5000$ and $d=2$, and consider the two different choices of assessment function discussed in Section \ref{sec:dlsn_state_est}. For $B=1$, we take $\nu_{\tau_{t,s}}(\bm{U}) = p(\mathcal{Y}_{t+1} | \bm{U})$, and for $B \in \{2,3\}$ we take the assessment function given by \eqref{eq:eta_uwithB}. For the online procedure, the results are summarised in Figures \ref{fig:online_cases_ess_mse}, \ref{fig:online_cases_rocs} and \ref{fig:theta_online_cases} and corresponding figures for the offline procedure can be found in \ref{app:altscen}. To assess the performance Figure \ref{fig:online_cases_ess_mse} depicts the ESS, mean square error in probability, given by
\begin{align}
MSE_{prob}^{(t)} = \dfrac{1}{{N \choose 2}} \sum_{i<j} \left(p_{ijt} - \hat{p}_{ijt}\right)^2, \label{eq:mse_prob}
\end{align}
and Figure \ref{fig:online_cases_rocs} shows ROC curves. Figure \ref{fig:theta_online_cases} shows the parameter estimates.

Overall, we see that increasing the number of look-ahead steps $B$ does not offer much improvement in terms of the performance of the filter. This is likely due to the observations being binary, and setting $B>1$ may offer improvement for more informative observations. The cases \ref{item:type1} and \ref{item:type2} exhibit the most stable performance throughout and this is particularly clear in Figure \ref{fig:theta_online_cases} where we see that the estimates of $\alpha$ and $\sigma$ for case \ref{item:type3} vary according to the decreasing density of the observed networks. We also see in Figure \ref{fig:cases_online_mse} that the MSE in probability is larger for case \ref{item:type3}, however this is somewhat obscured by the variability in the parameter estimates. In the offline setting, we find that the parameter estimates are more stable which results in a larger MSE (see \ref{app:altscen}). Online estimation is challenging in general, and so it is unsurprising that we find that the performance is generally much more stable in the offline setting. Finally, we comment that suitable modifications of our model may be made in order to express the structures present in \ref{item:type2} and \ref{item:type3} if appropriate. 

\subsection{Scalability}
\label{sec:sims_scalability}

In this section we explore the scalability of our approach as the dimension of the state space increases in terms $N$ and $T$. We simulate data according to the model in Section \ref{sec:dlsn_model} under the following two settings.
\begin{enumerate}
  \item \textbf{Increasing $T$}: Simulate a dataset with $T=1000$ and $N=20, d=2, \alpha=1.25, \sigma=0.2,$ $\phi=0.9$, and restrict the number of observations to $t=50, 100, 500,$ and $1000$.
  \item \textbf{Increasing $N$}: Simulate a dataset for each of $N \in \{50, 75, 100\}$ and $T=25, d=2,$ $\alpha=1, \sigma=0.2, \phi=0.9$.
\end{enumerate}

\begin{figure}[t]
  \centering 
    \begin{subfigure}[t]{.45\textwidth}
        \centering
        \includegraphics[trim={0 0.5cm 0cm 1.5cm}, clip, height=.6\textwidth]{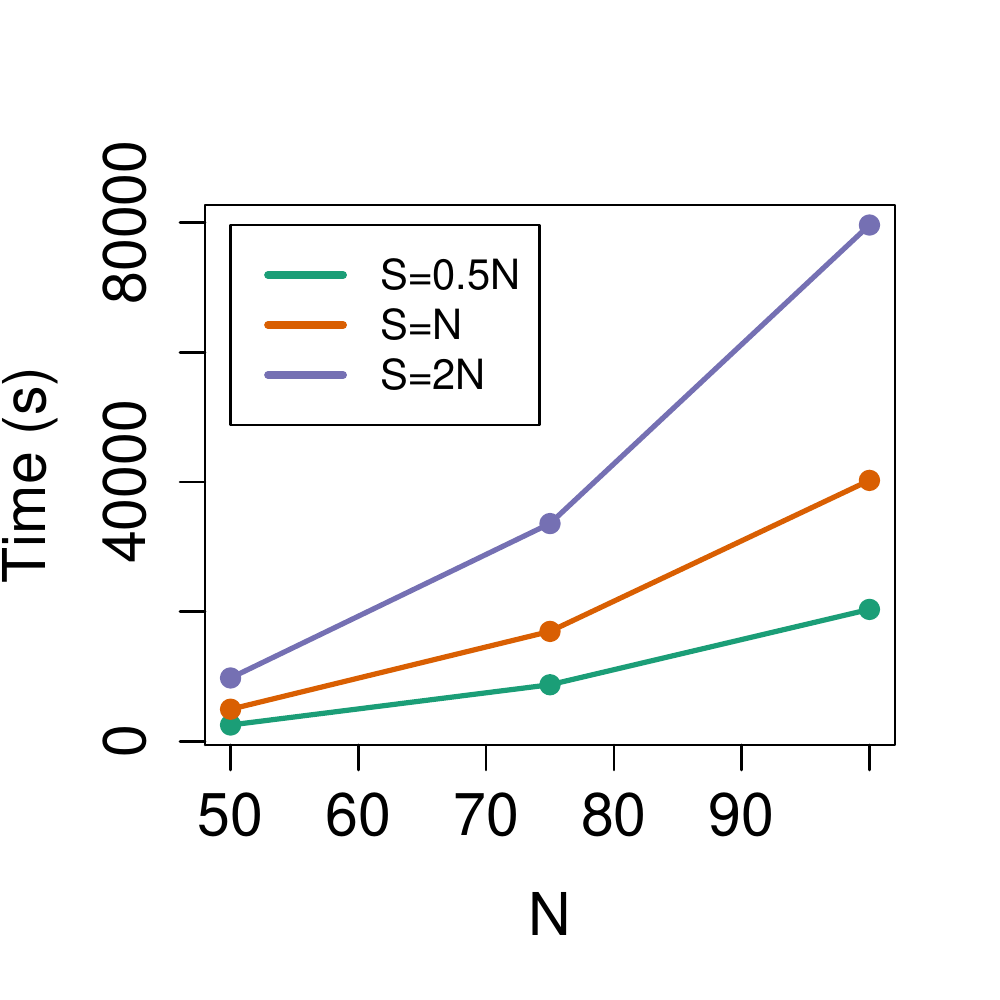}
        \caption{Computational time as a function of $N$ for $T=25$. The number of intermediary states is set to $0.5N, N$ and $2N$.}
        \label{fig:incNscl}
    \end{subfigure}
    \hspace{.5cm}
    \begin{subfigure}[t]{.45\textwidth}
      \centering
      \includegraphics[trim={0cm 0.5cm 0 1.5cm}, clip, height=.6\textwidth]{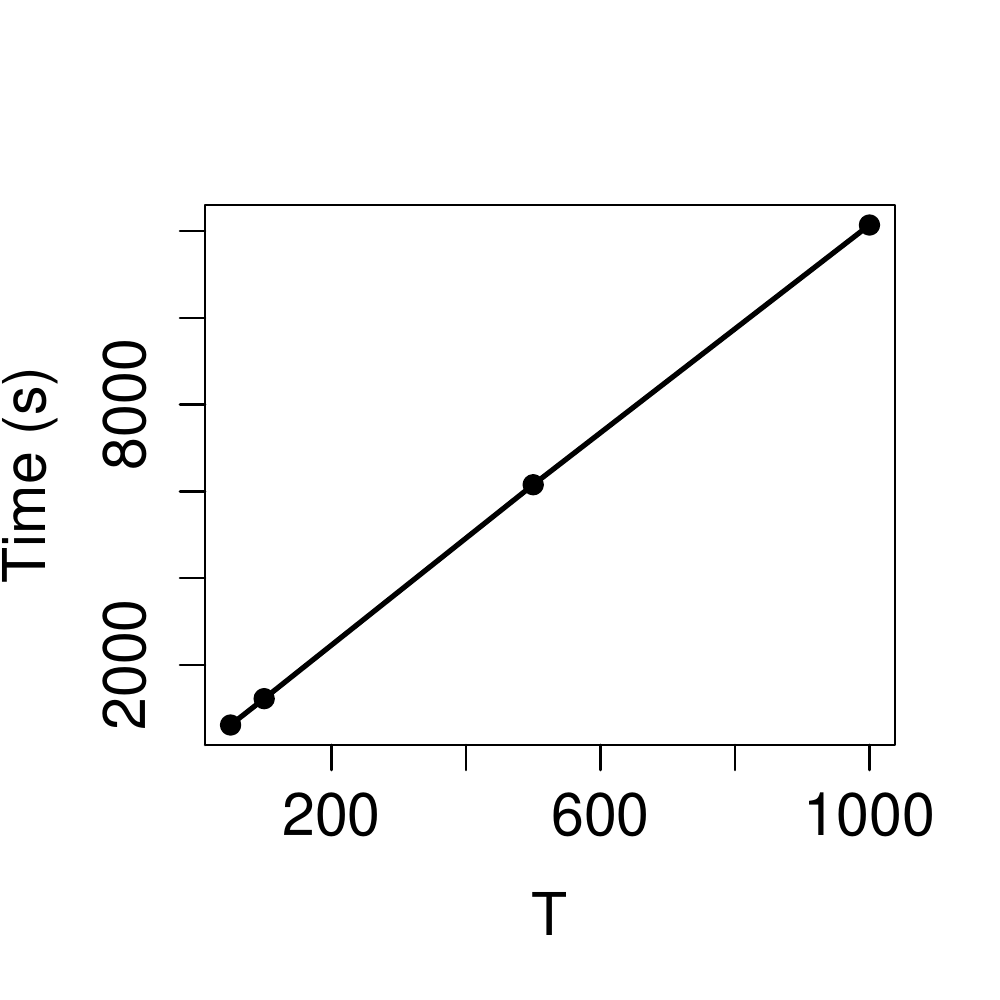}
      \caption{Time as a function of $T$ for $N=20$. The number of intermediary states is set to $S=N$.}
      \label{fig:incTscl}
    \end{subfigure}
  \caption{Time taken for online SMC procedure for increasing $N$ (left) with $S=0.5N$ (green), $S=N$ (orange) and $S=2N$ (purple), and increasing $T$ (right).} \label{fig:scale_times}
\end{figure}

For each case we implement online inference with $M=5000$ and the timings are shown in Figure \ref{fig:scale_times}. We choose the number of intermediary states as $S=N$ for increasing $T$ and $S=0.5N, N,$ and $2N$ for increasing $N$. From Figure \ref{fig:incNscl} we see that the computational cost grows quadratically with the number of nodes $N$ and this is due to the ${N \choose 2}$ terms in the expression \eqref{eq:pygivenu}. Whilst we may reduce the overall computational cost through decreasing $S$, this will affect the performance of the filter (see Figure \ref{fig:scale_summ_Ninc} in \ref{app:scale_perf}). From Figure \ref{fig:incTscl} we see that there is a linear increase in the computational cost as $T$ grows. The filter is much more stable in this setting (see Figure \ref{fig:scale_summ_Tinc} in \ref{app:scale_perf}), and so is much more appropriate for network data with a large number of observations in time.

\section{Classroom Contact Data}
\label{sec:realdat}
 
In this section we consider the application of our proposed methodology to a dataset describing face-to-face contact among primary school children. Students are equipped with sensors and a connection is recorded if two students face each other within a 20 second interval. The data are available from \texttt{www.sociopatterns.org} and have been published in \cite{stehle2011} and \cite{gemmetto2014}. We analyse interactions among a class of 25 school children on an aggregate level, where we record whether each pair of students interacted within a 4 minute interval. In Section \ref{sec:clsrm_binary} we analyse the data with binary interactions whereas in Section \ref{sec:clsrm_integer} we analyse the data with integer interactions. 

\subsection{Binary Interactions}
\label{sec:clsrm_binary}

We now apply our model to the contact network where the connections are binary, so that $y_{ijt}=1$ if students $i$ and $j$ interact within the $t^{th}$ interval and $y_{ijt}=0$ otherwise. We compare our model to the approach of \cite{durante2014} who model the latent trajectories according to a Gaussian process and, throughout this section, we report estimates obtained via the offline procedure. In addition to the formulation presented in Section \ref{sec:dlsn_model}, we also fit our model with the linear predictor \eqref{eq:m_eta} replaced by 
\begin{align}
  \eta_{ijt} = \alpha + u_{it}^T u_{jt} \label{eq:m_eta_dp}.
\end{align}
Similarly to the Euclidean distance formulation, $u_{it}$ captures the probability of the $i^{th}$ node interacting with the remaining nodes at time $t$ relative to the base-rate parameter $\alpha$. In contrast to the Euclidean distance assumption, the dot-product assumption imposes that nodes whose latent positions lie in similar directions from the origin are more likely to be connected. 

Since \cite{durante2014} model the connection probabilities as a function of the dot-product between latent variables, the formulation \eqref{eq:m_eta_dp} makes for a more reasonable comparison than the Euclidean distance formulation in presented in Section \ref{sec:dlsn_model}. We note that these models differ in how the latent trajectories and the base-rate parameter $\alpha$ are modelled. More specifically, \cite{durante2014} assume that the latent coordinates evolve according to a Gaussian process and allow $\alpha$ to vary with time. 

\begin{figure}[t]
  \centering
  \begin{subfigure}[b]{\textwidth}
  \includegraphics[trim={0cm, 0.2cm, 0.4cm, 1cm}, clip, width=\textwidth]{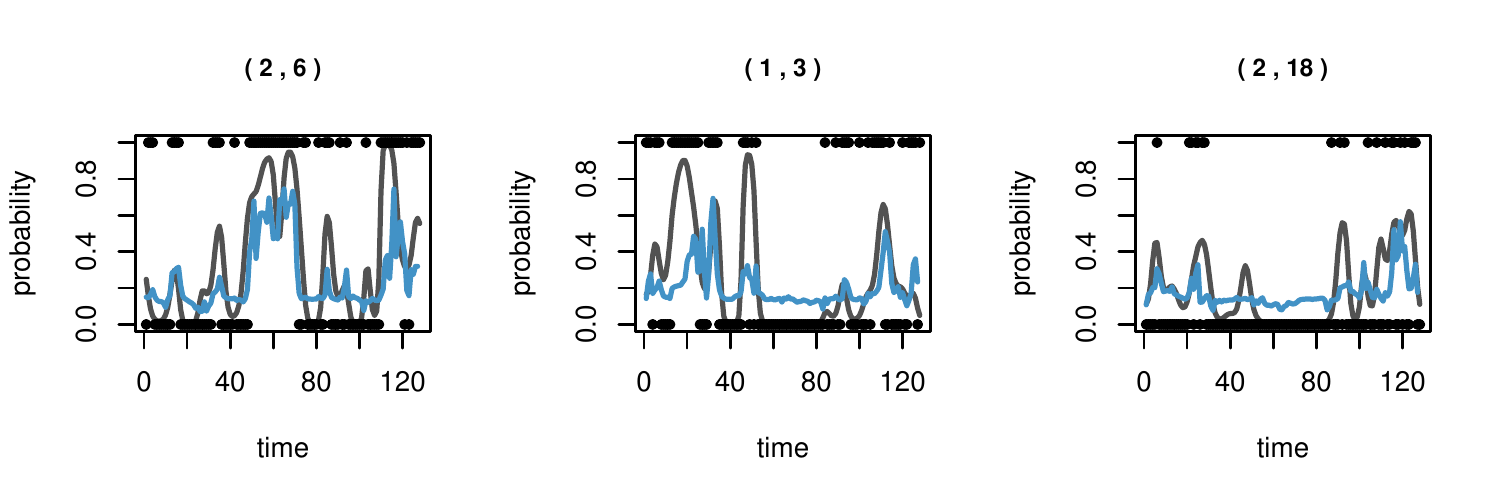}
  \caption{Posterior mean connection probabilities for the dot-product DLSN model (blue) and the GP model of \cite{durante2014} (grey).} \label{fig:prim_prob_bin}
  \end{subfigure} 
  \begin{subfigure}[t]{\textwidth}
    \includegraphics[trim={0cm, 0.2cm, 0.4cm, 1cm}, clip, width=\textwidth]{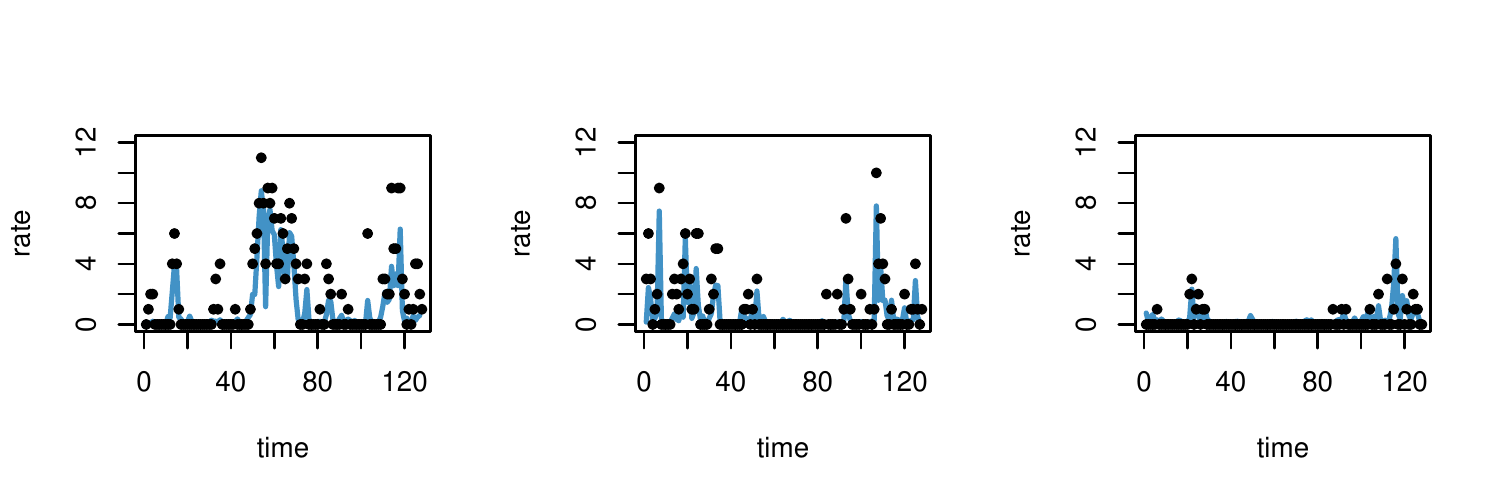}
  \caption{Connection rates for the dot-product DLSN model.} \label{fig:prim_rate_wght}
  \end{subfigure}
  \caption{Fitted model obtained via offline procedure for a selection of student pairs. Figure \ref{fig:prim_prob_bin} shows the connection probabilities for binary data and Figure \ref{fig:prim_rate_wght} shows the rate for count data. Left, middle and right correspond to the same node pairs for each plot.}
\end{figure} 

We fit our model with $S=2N, M=5000$ and $d=2$ and, similarly to \cite{durante2014}, we assess the model on the connection probabilities to avoid identifiability issues associated with the latent representation. Figure \ref{fig:prim_prob_bin} shows the estimated connection probabilities for the first $T-1$ observations for a selection of node pairs, where the blue line represents the dot-product DLSN model fitted and the grey line represents the model of \cite{durante2014}. We see that there is a reasonable correspondence between the two models, however the model of \cite{durante2014} returns a smoother estimate of the probability trajectory. Our approach instead is able to obtain estimates at a lower computational cost, particularly when online inference is implemented. More precisely, the MCMC scheme of \cite{durante2014} involves update steps for each time point and node at every interaction, whereas our online GIRF scheme requires $S \times M$ likelihood evaluations at each time point.

\begin{figure}[t]
  \centering
  \begin{subfigure}[t]{.31\textwidth}
  \centering
  \includegraphics[trim={0cm, 0cm, 0cm, 0.5cm}, clip, width=\textwidth]{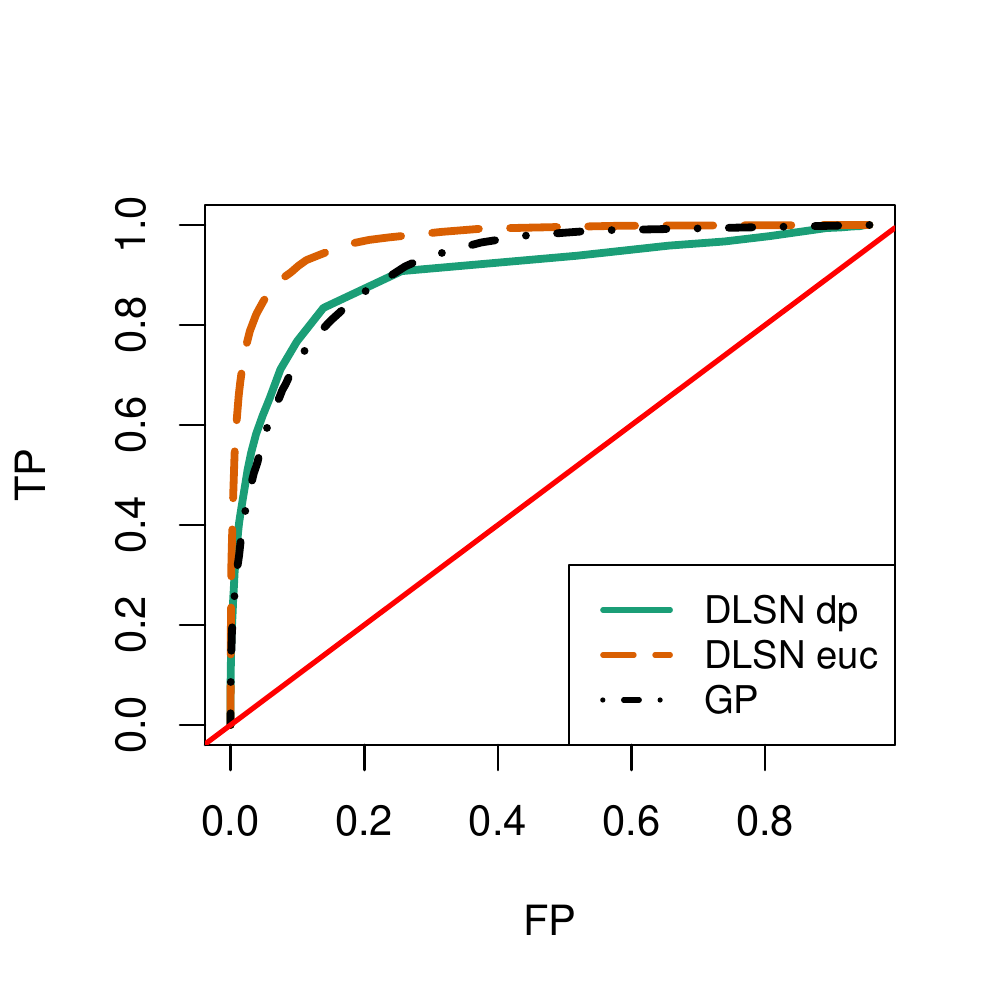}
  \caption{ROC curves.}
  \label{fig:prim_roc}
  \end{subfigure}
  ~
  \begin{subfigure}[t]{.31\textwidth}
    \centering
  \includegraphics[trim={0cm, 0cm, 0cm, 0.5cm}, clip, width=\textwidth]{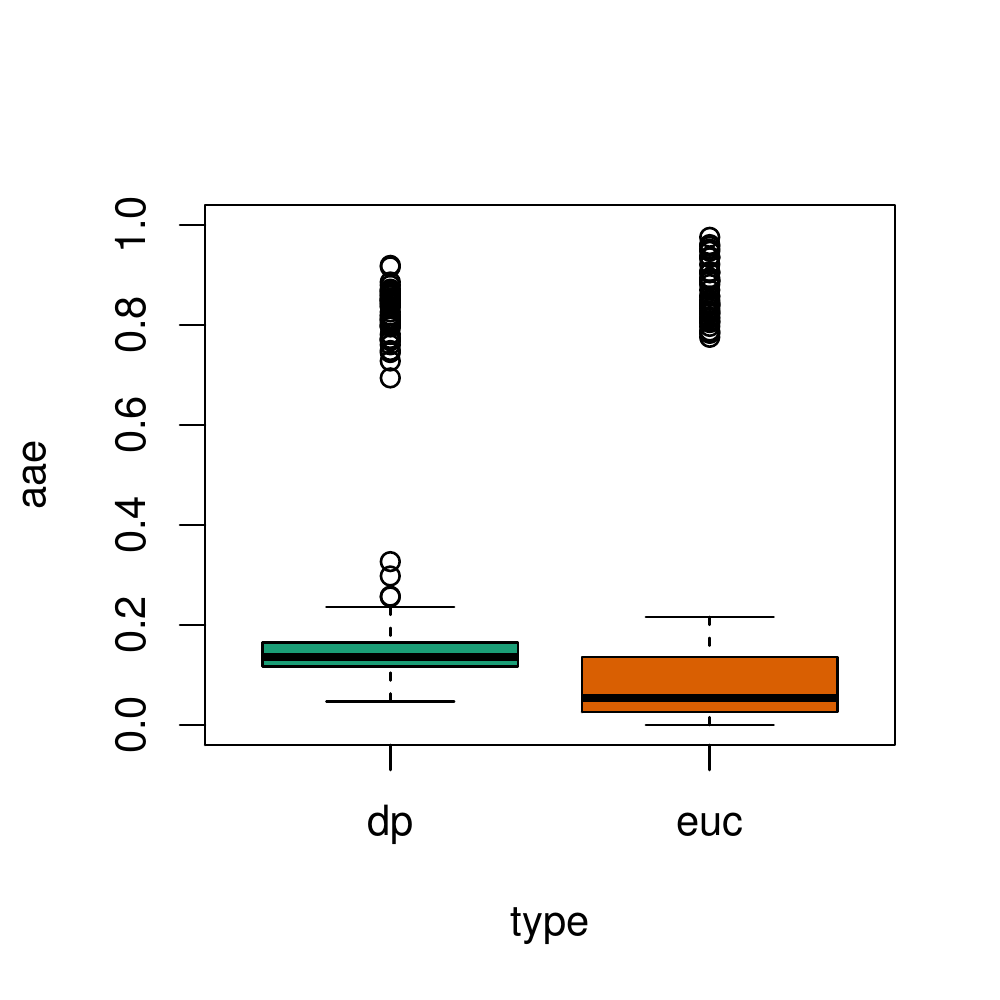}
  \caption{AAE for predictive probabilities.}
  \label{fig:prim_bin_pred}
  \end{subfigure}
  ~
  \begin{subfigure}[t]{.31\textwidth}
    \centering
  \includegraphics[trim={0cm, 0cm, 0cm, 0.5cm}, clip, width=\textwidth]{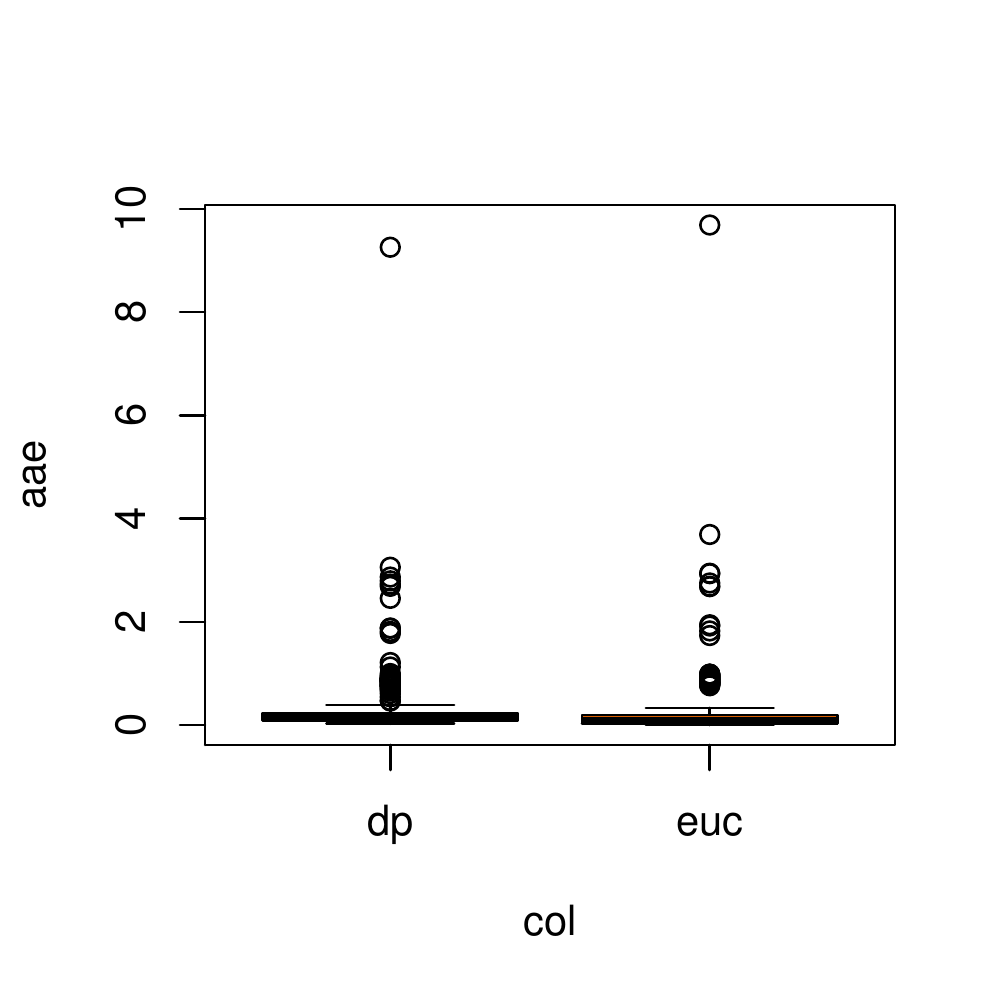}
  \caption{AAE for predictive rates.}
  \label{fig:prim_w_pred}
  \end{subfigure}
  \caption{Summary of fit for the dot-product DLSN, Euclidean distance DLSN and the model of \cite{durante2014}. Figure \ref{fig:prim_roc} shows the ROC curve for each model. Figures \ref{fig:prim_bin_pred} and \ref{fig:prim_w_pred} compare the average absolute error (AAE) \eqref{eq:aae_prob} for our model with a dot-product and Euclidean distance formulations when the interactions are binary and weighted, respectively. We note the predictive measures are only considered for the DLSN models.}
  \label{fig:prim_bin_fit}
\end{figure}

Figure \ref{fig:prim_roc} reports the ROC curve for the DLSN model and the model of \cite{durante2014}, and the dot-product and Euclidean distance formulations of the DLSN model have both been included for comparison. We see from this figure that all the models perform comparably well in terms of this measure, however we see that the Euclidean distance model is a marginally better classifier. 

Finally, we explore the predictive distribution for the $T^{th}$ observation. To assess the quality of predictions, we simulate $R_{rep}=5000$ observations according to $\hat{p}_{ijT}$ and record the average absolute error. For $\{i,j\} \in \{i,j \in \{1,2,\dots,N\} | i<j\}$, this is given by
\begin{align}
  AAE_{ij} = \dfrac{1}{R_{rep}} \sum_{r = 1}^{R_{rep}} \left| y_{ijT} - \hat{y}_{ijT}^r \right|, \label{eq:aae_prob}
\end{align}
where $\hat{y}_{ijT}^r$ is simulated from a Bernoulli distribution with success probability $\hat{p}_{ijT}$. Note that \eqref{eq:aae_prob} is equal to 0 when the data are predicted correctly. The distribution of the AAE over all node pairs is depicted in Figure \ref{fig:prim_bin_pred} for the dot-product and Euclidean distance DLSN model, and from this we see that most observations are well predicted. Additionally, we observe that the Euclidean distance formulation performs better in terms of this measure.

\subsection{Weighted Interactions}
\label{sec:clsrm_integer}

Our approach also allows us to straightforwardly analyse networks with non-binary interactions and, in this section, we consider the classroom data where $y_{ijt}$ represents the number of interactions between nodes $i$ and $j$ in the $t^{th}$ time interval. To adapt the model in Section \ref{sec:dlsn_model}, we model the probability of connections forming as
\begin{align}
    p( \mathcal{Y}_t | \bm{U}_t, \alpha ) &=  \prod_{i < j} e^{-\lambda_{ijt}} \lambda_{ijt}^{y_{ijt}} / y_{ijt}!, \label{eq:pygivenu_poiss} \\
  \lambda_{ijt} &= \exp\{ \alpha - \|u_{it} - u_{jt} \| \}, \label{eq:modellink_poiss} 
\end{align}
with expressions for the latent transitions \eqref{eq:u0} and \eqref{eq:ut_transition} specified as before.

The formulation \eqref{eq:modellink_poiss} can also be adapted to express $\lambda_{ijt}$ in terms of the dot-product $u_i^T u_j$, similar to the model considered in Section \ref{sec:clsrm_binary}. Here, we consider both formulations and explore the model fit when estimates are obtained via the offline procedure. Throughout, we fix $S=2N, M=5000$ and $d=2$. The estimated rates for a subset of node pairs for the dot-product formulation is shown in Figure \ref{fig:prim_rate_wght}, where each panel corresponds to the same set of node pairs in Figure \ref{fig:prim_prob_bin}. It is clear that modelling the data in this way allows us to obtain an understanding of the interactions on a finer-scale. Finally, we assess the predictive distribution for the $T^{th}$ observations in Figure \ref{fig:prim_w_pred}. We see that the majority of observations are predicted well and that there is no clear difference between the dot-produce and Euclidean distance models. Note that we have not compared to the approach of \cite{durante2014} for this example. Since their inference procedure relies on a data augmentation scheme for logistic models, it is not immediately straightforward to adapt their approach to the non-binary setting.

\section{Discussion}
\label{sec:smc_disc}

In this article we have considered sequential estimation of a dynamic latent space network model and our approach relies on the algorithms introduced in \cite{park2019} and \cite{nemeth2016}. Since standard SMC methodology scales poorly as the dimension of the state space increases, we consider the GIRF of \cite{park2019} which allows us to estimate networks with moderate $N$. Our approach is most appropriate in settings where there are a large number of observations in time, allows convenient estimation of predictive distributions and, unlike MCMC, can be updated with additional observations without rerunning the entire inference mechanism. We have also shown that, in contrast to related approaches, our procedure can be adapted to model variations, such as weighted interactions, and is suited to both online and offline inference. To the best of our knowledge, we have presented a novel approach for sequential estimation of dynamic latent space networks which, in contrast to existing methodology, does not rely on specific model forms or approximations.

A key limitation of our approach is the scalability in terms of the number of nodes. This is due to both properties of SMC methodology and also the $O(N^2)$ calculations needed to evaluate the likelihood. The latter has been addressed via approximations (see \cite{raftery2012} and \cite{rastelli2018}), though it is not straightforward to adapt this approach within the context of SMC.  Alternatively, the modelling approach taken in \cite{fosdick2016}, in which the nodes are partitioned into communities and the within-community connection probabilities are modelled via a latent space, may allow us to consider networks with larger $N$. In this setting, regions of independence in the latent space may allow us to develop a more scalable approach by partitioning the state space (for example, as in \cite{rebeschini2013}). 

This work may also be considered in the context of changepoint or anomaly detection, similarly to \cite{lee2019}, where the authors rely on approximate inference via variational methods. However, in contrast to existing DLSN models, our approach models the latent nodes via a stationary process and so is well suited to finding changes in network behaviour. As an example, the latent space approach may allow for the detection of changes in joint distributions of motif counts resulting from a change in the variance of the latent representation. 

\section*{Acknowledgements}

KT gratefully acknowledges the support of the EPSRC funded  EP/L015692/1 STOR-i Centre for Doctoral Training and CN gratefully acknowledges the support of EPSRC grants EP/S00159X/1, EP/R01860X/1 and EP/V022636/1. We thank Daniele Durante for providing code for the Gaussian Process network model used in Section \ref{sec:clsrm_binary}, and we thank Brendan Murphy and Marco Battiston for helpful comments on an earlier version of this work. \\

Code will be made available providing details of the implementations for this paper.

\setlength{\bibsep}{0pt plus 0.3ex}
\bibliographystyle{apalike}
\bibliography{biblio}

\appendix

\section{Derivation of $\tilde{p}_{\theta}(\bm{U}_{\tau_{t,s}} | \bm{U}_{\tau_{t,s-1}})$}
\label{app:trans_eq_ptilde}

In this appendix we derive the intermediary transition kernels required to implement the GIRF. Since all nodes follow the same distribution independently, we drop the notation $i$ and derive the transition equation for a single node. 

For $u_{t}$ we have $U_{t} | U_{t-1}=u_{t-1} \sim \mathcal{N}(\phi u_{t-1}, \sigma^2 I_d)$. We require an expression for $\tilde{p}_{\theta}(u_{\tau_{t,s}} | u_{\tau_{t,s-1}})$ so that
\begin{align}
  p_{\theta}(u_{t+1} | u_{t}) = \tilde{p}_{\theta}(u_{\tau_{t,1}} | u_{t}) \tilde{p}_{\theta}(u_{\tau_{t,2}} | u_{\tau_{t,1}}) \dots \tilde{p}_{\theta}(u_{t+1} | u_{\tau_{t,S-1}}).
\end{align}

By properties of the Normal distribution, we know that $\tilde{p}_{\theta}(u_{\tau_{t,s}} | u_{\tau_{t,s-1}}) = \mathcal{N}( m u_{\tau_{t,s-1}}, v I_d)$. We determine the values of $m$ and $v$ below. \\

\textbf{Determining $m$} \\
By the law of total expectation we have
\begin{align}
  \mathbb{E}\left[ U_{t} | U_{t-1} \right] &= \mathbb{E}\left[ U_{\tau_{t,S}} | U_{\tau_{t,0}} \right] = \phi u_{t-1}\\
                                           &= \mathbb{E}\left[ \mathbb{E} \left[\dots \mathbb{E} \left[ \mathbb{E} \left[ U_{\tau_{t,S}} | U_{\tau_{t,S-1}} \right] \vert U_{\tau_{t,S-2}} \right] \dots | U_{\tau_{t,1}} \right]\vert U_{\tau_{t,0}} \right] \\
  &= m^S u_{t-1} \\
  \Rightarrow m &= \phi^{1/S} 
\end{align}
We have $\tilde{p}_{\theta}(u_{\tau_{t,s}} | u_{\tau_{t,s-1}}) = \mathcal{N}( \phi^{1/S} u_{\tau_{t,s-1}}, v I_d)$ and we note that this is valid for $\phi \in (0,1)$. \\

\textbf{Determining $v$} \\
The law of total variance tells us
\begin{align}
\Var\left[  U_{\tau_{t,S}} | U_{\tau_{t,0}} \right] &= \mathbb{E} \left[ \Var \left[ U_{\tau_{t,S}} | U_{\tau_{t,1}} \right] | U_{\tau_{t,0}} \right] + \Var \left[ \mathbb{E} \left[ U_{\tau_{t,S}} | U_{\tau_{t,1}} \right] | U_{\tau_{t,0}} \right] \label{eq:var_uS_u0}
\end{align}
For the expectation, we have
\begin{align}
  \mathbb{E} \left[ U_{\tau_{t,s}} | U_{\tau_{t,s'}} \right] = \phi^{(s - s')/S} u_{\tau_{t,s}} \mbox{ for } s' < s.
\end{align}
We then obtain a recursion for \eqref{eq:var_uS_u0} by writing
\begin{align}
\Var\left[  U_{\tau_{t,S}} | U_{\tau_{t,s}} \right] &= \mathbb{E} \left[ \Var \left[ U_{\tau_{t,S}} | U_{\tau_{t,s+1}} \right] | U_{\tau_{t,s}} \right] + \Var \left[ \mathbb{E} \left[ U_{\tau_{t,S}} | U_{\tau_{t,s+1}} \right] | U_{\tau_{t,s}} \right]
\end{align}
for $s = S-2, S-3, \dots, 0$. \\

Starting from $s=S-2$, we have
\begin{align}
\Var\left[  U_{\tau_{t,S}} | U_{\tau_{t,S-2}} \right] &= \mathbb{E} \left[ \Var \left[ U_{\tau_{t,S}} | U_{\tau_{t,S-1}} \right] | U_{\tau_{t,S-2}} \right] + \Var \left[ \mathbb{E} \left[ U_{\tau_{t,S}} | U_{\tau_{t,S-1}} \right] | U_{\tau_{t,S-2}} \right]   \\
  &= \mathbb{E} \left[ v I_d  \vert U_{\tau_{t,S-2}} \right] + \Var \left[ \phi^{1/S} U_{\tau_{t,S-1}} \vert U_{\tau_{t,S-2}} \right] \\
  &= vI_d + \phi^{2/S}  v I_d
\end{align}
Then from $s = S-3$ we have
\begin{align}
  \Var\left[  U_{\tau_{t,S}} | U_{\tau_{t,S-3}} \right] &= \mathbb{E} \left[ \Var \left[ U_{\tau_{t,S}} | U_{\tau_{t,S-2}} \right] | U_{\tau_{t,S-3}} \right] + \Var \left[ \mathbb{E} \left[ U_{\tau_{t,S}} | U_{\tau_{t,S-2}} \right] | U_{\tau_{t,S-3}} \right] \\
                                                        &= \mathbb{E} \left[ vI_d + \phi^{2/S} vI_d \vert U_{\tau_{t,S-3}} \right] + \Var \left[ \phi^{2/S} U_{\tau_{t,S-2}} | U_{\tau_{t,S-3}} \right] \\
                                                        &= v I_d + \phi^{2/S} v I_d + \phi^{4/S} v I_d
\end{align}
Applying this procedure iteratively, we obtain
\begin{align}
  \sigma^2 = \Var\left[  U_{\tau_{t,S}} | U_{\tau_{t,0}} \right] &= v I_d\left[ 1 + \phi^{2/S} + \phi^{4/S} + \dots + \phi^{2(S-1)/S} \right] \\
                                                                 &= v I_d \sum_{r=0}^{S-1} \left(\phi^{2/S}\right)^r = v I_d \left( \dfrac{1 - \left(\phi^{2/S}\right)^S }{1 - \phi^{2/S} } \right) = v I_d \left( \dfrac{1 - \phi^{2} }{1 - \phi^{2/S} } \right) \\
\Rightarrow v &= \sigma^2 \left( \dfrac{1 - \phi^{2/S}}{1 - \phi^2} \right) I_d
\end{align}

Finally, we have
\begin{align}
  { \tilde{p}_{\theta}(u_{\tau_{t,s}} | u_{\tau_{t,s-1}}) = \mathcal{N}\left( \phi^{1/S} u_{\tau_{t,s-1}}, \sigma^2 \left( \dfrac{1 - \phi^{2/S}}{1 - \phi^2} \right)I_d \right) }
\end{align}


\section{Gradient derivation for parameter estimation}
\label{app:theta_gradients_alt} 

To estimate $\theta = (\alpha, \sigma, \phi)$ using the scheme detailed in Section \ref{sec:dlsn_est}, we must find expressions for $\dfrac{d}{d \theta} \log p(\bm{U}_t | \bm{U}_{t-1}, \theta)$ and $\dfrac{d}{d \theta} \log p(\mathcal{Y}_t | \bm{U}_{t}, \theta)$. For the model detailed in Section \ref{sec:dlsn_model} we obtain
\begin{align}
  \dfrac{\partial}{\partial \alpha} \log p(\mathcal{Y}_t | \bm{U}_{t}, \theta) &= \sum_{i<j} \left\{ y_{ijt} - \dfrac{1}{1 + \exp(\| u_{it} - u_{jt} \| - \alpha ) } \right\} \\
  &= \sum_{i<j} \left( y_{ijt} - p_{ijt} \right) \\
  \dfrac{\partial}{\partial \sigma} \log p(\mathcal{Y}_t | \bm{U}_{t}, \theta) &= \dfrac{\partial}{\partial \phi} \log p(\mathcal{Y}_t | \bm{U}_{t}, \theta) = \dfrac{\partial}{\partial \alpha} \log p(\bm{U}_t | \bm{U}_{t-1}, \theta) = 0 \\
  \dfrac{\partial}{\partial \sigma} \log p(\bm{U}_t | \bm{U}_{t-1}, \theta) &= \sum_{i=1}^N \left\{  -\dfrac{d}{\sigma} + \dfrac{1}{\sigma^3} (u_{it} - \phi u_{i,t-1} )^T (u_{it} - \phi u_{i,t-1} ) \right\} \label{eq:ddsig} \\
  \dfrac{\partial}{\partial \phi} \log p(\bm{U}_t | \bm{U}_{t-1}, \theta) &= \sum_{i=1}^N \left\{ \dfrac{1}{\sigma^2} u_{i,t-1}^T (u_{it} - \phi u_{i,t-1} ) \right\} \label{eq:ddphi} 
\end{align}
where \eqref{eq:ddphi} follows from (84) in \cite{peterson2012}. 

Since $\sigma >0$ and $\phi \in (0,1)$, we opt to estimate $\tilde{\sigma}$ and $\tilde{\phi}$ as specified in \eqref{eq:tilde_sigandphi}. From the chain rule we have
\begin{align}
  \dfrac{\partial}{\partial \tilde{\sigma}} = \dfrac{\partial}{\partial \sigma} \dfrac{\partial \sigma}{\partial \tilde{\sigma}} \hspace{1cm} \mbox{and} \hspace{1cm} \dfrac{\partial}{\partial \tilde{\phi}} = \dfrac{\partial}{\partial \phi} \dfrac{\partial \phi}{\partial \tilde{\phi}}.
\end{align}
Then, we obtain the required expressions using
\begin{align}
  \dfrac{\partial \tilde{\sigma}}{\partial {\sigma}} = \dfrac{1}{\sigma}, \hspace{1cm} \mbox{and} \hspace{1cm} \dfrac{\partial \tilde{\phi}}{\partial {\phi}} = \dfrac{1}{\phi (1 - \phi)}.\\
   %
\end{align}

For stability reasons, we scale each of the gradients by a $1/Nd$.

\section{Suitability of the GIRF}
\label{app:check_girf}

\begin{figure}[t!]
    \centering
    \begin{subfigure}[t]{.45\textwidth}
        \centering
        \includegraphics[width=\textwidth]{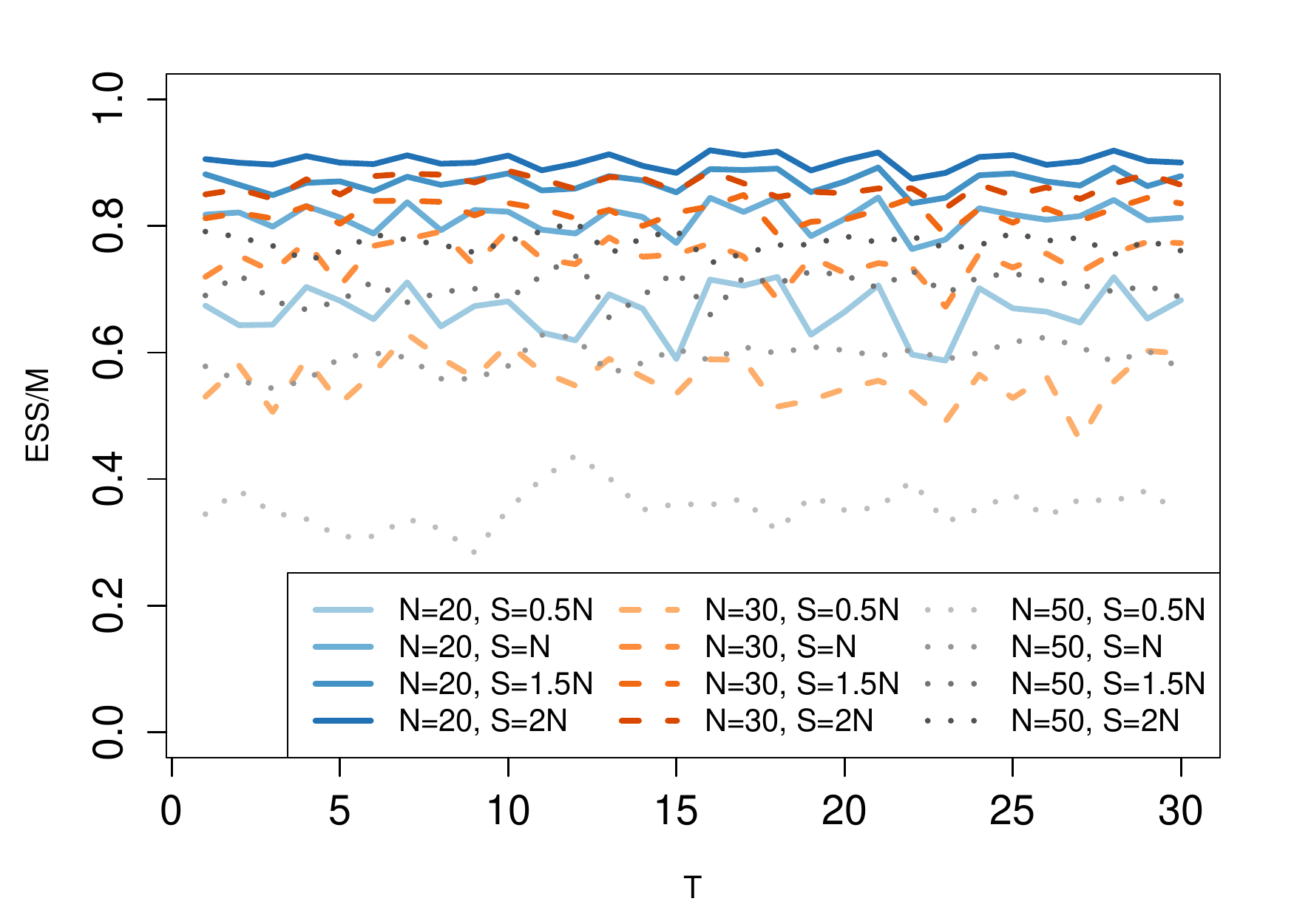}
    \end{subfigure}
    \begin{subfigure}[t]{.45\textwidth}
        \centering
        \includegraphics[width=\textwidth]{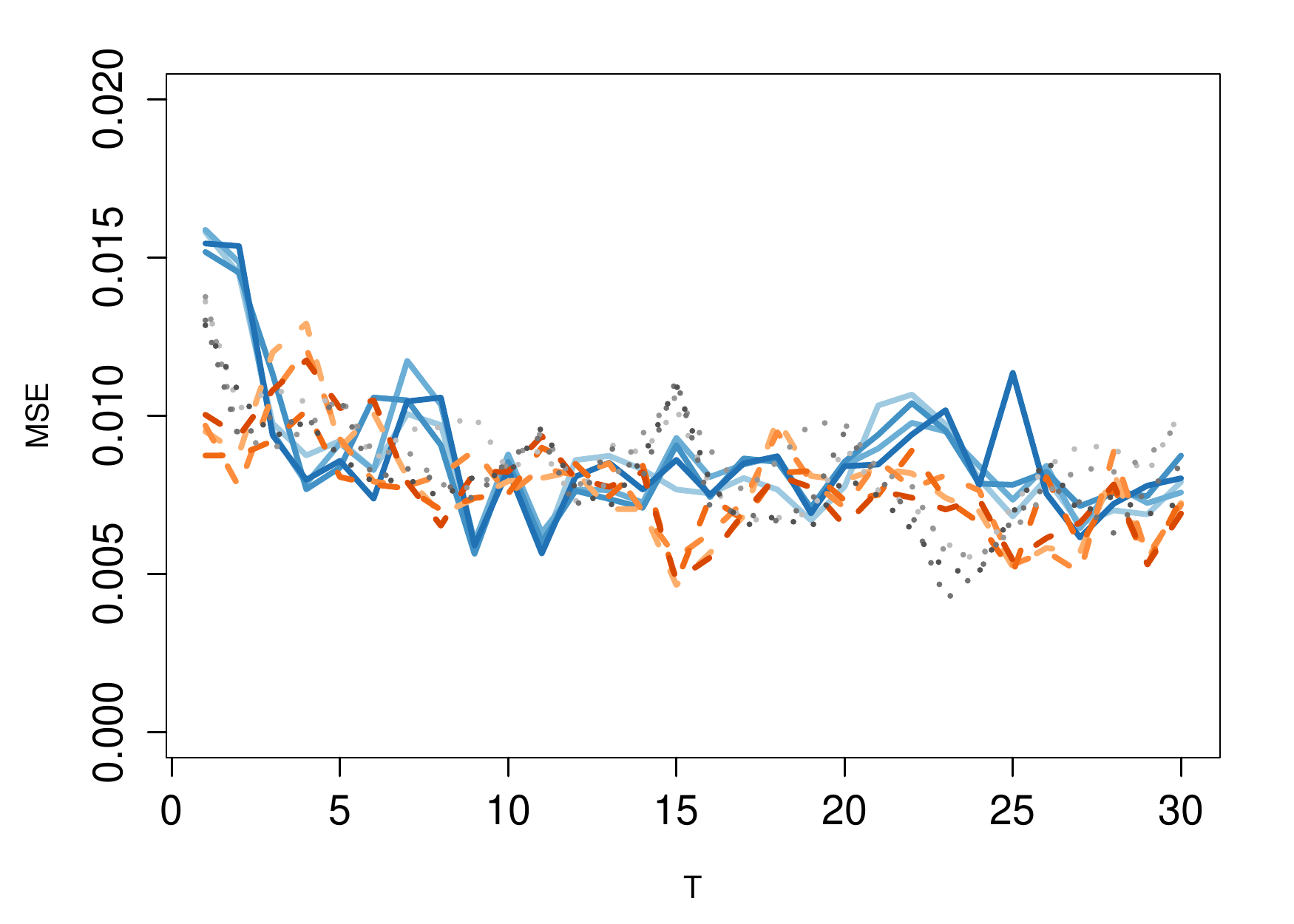}
    \end{subfigure}
    \caption{Performance of GIRF as $N$ increases for varying number of intermediary states $S$. The ESS and average MSE in probability are shown in the left and right panels, respectively, and the number of particles was fixed at $M=10000$. Blue, red and grey lines correspond to $N=20$, $N=30$ and $N=50$, respectively.}
    \label{fig:girf}
\end{figure}

To assess the suitability of this approach, we fit the GIRF to data simulated from the model in Section \ref{sec:dlsn_model} with $\alpha = 1.2, \sigma = 0.2$ and $\phi = 0.9$. Figure \ref{fig:girf} shows the performance of the GIRF for different choices of $S$ and varying $N$ when the assessment function is taken as $v_{\tau_{t,s}}(\bm{U}) = p(\mathcal{Y}_{t+1} | \bm{U})$ and $\theta$ is assumed known. The left plot depicts the ESS (see Section \ref{sec:pf_ssm}) divided by the number of particles and the right hand size shows the mean square error in probability. For $t=1,2,\dots,T$, this is given by
\begin{align}
  MSE_{prob}^{(t)} = \dfrac{1}{{N \choose 2}} \sum_{i<j} \left(p_{ijt} - \hat{p}_{ijt}\right)^2, 
\end{align}
where $p_{ijt}$ and $\hat{p}_{ijt}$ represent the true and estimated probability of nodes $i$ and $j$ sharing a tie at time $t$, respectively. 

Figure \ref{fig:girf} indicates that the accuracy of the filter remains reasonably stable as the dimension of the state space increases. We also see that estimation is more challenging as $N$ increases and that $S$ significantly impacts the performance of the filter. It is clear that too few intermediary states will result in poor quality approximations. Finally, we comment here that little difference in the performance was observed when the assessment function was given by \eqref{eq:uwithB} and it is not clear whether this is true more generally. For example, we may find that incorporating future observations is more worthwhile for non-binary interactions.

\section{Static parameter initialisation}
\label{app:theta_init}

To initialise $\sigma$ we follow \cite{sewell2015} and adapt the GMDS approach considered in \cite{sarkar2006} and our procedure is detailed in Algorithm \ref{alg:siginit}. We initialise $\phi$ at $0.8$, and then choose $\alpha$ via a grid search conditional on $\sigma$ and $\phi$. More specifically, for a sequence of candidate $\alpha$ values, we simulate a series of latent trajectories given $\sigma$ and $\phi$. Then, we initialise $\alpha$ by taking the value which results in a network with density that is most similar to the average observed density ${\sum_{t=1}^T \sum_{i < j} y_{ijt} }/{T{N \choose 2}}$.

\begin{algorithm}[t]
  \caption{Procedure for initialising $\sigma$} \label{alg:siginit}
  \begin{algorithmic}
    \STATE \textbf{Input}: $\mathcal{Y}_1$, $\mathcal{Y}_2$ and $d$
    \STATE \textbf{For} $t \in \{1,2\}$:
    \STATE \hspace{1cm} - Calculate the $N \times N$ distance matrix $D_{t} = \{d_{ijt} \}_{i,j \in 1:N}$, where $d_{ijt}$ 
    \STATE \hspace{1.2cm} is the graph distance between nodes $i$ and $j$. We set $d_{iit}=0$ and, 
    \STATE \hspace{1.2cm} if $d_{ijt} = \infty$, we set $d_{ijt}$ to be the maximum finite entry of $D_t$
    \STATE \hspace{1.2cm} plus some constant $\gamma > 0$.
    \STATE \hspace{1cm} - Calculate the MDS coordinates with latent space dimension $d$, $\hat{\bm{U}}_t$
    \STATE \hspace{1cm} - Calculate $\sigma_t = \dfrac{1}{Nd} \sum_{i=1}^N \hat{u}_{it}$
    \STATE Let $\sigma = (\sigma_1 + \sigma_2)/2$ 
  \end{algorithmic}
\end{algorithm}

\section{Alternative Scenarios}
\label{app:altscen}

\begin{figure}[t!]
  \begin{subfigure}[t]{\textwidth}
    \centering
  \includegraphics[trim={0 6.5cm 0 1cm}, clip, width=\textwidth]{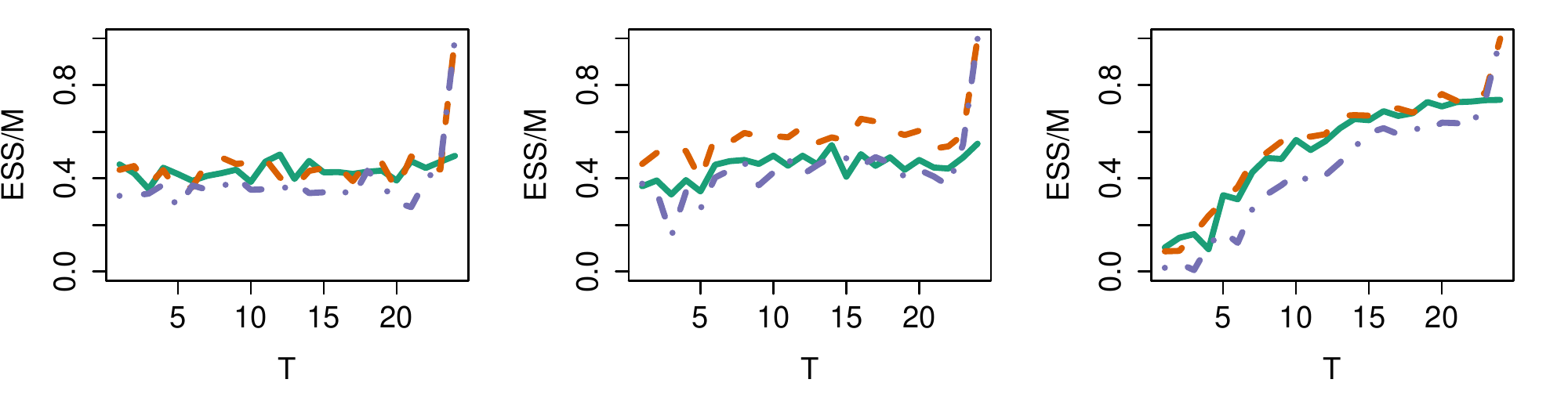}
  \caption{Effective sample size.} \label{fig:cases_offline_ess}
  \end{subfigure}
  \begin{subfigure}[t]{\textwidth}
    \centering
  \includegraphics[trim={0 0cm 0 7cm}, clip, width=\textwidth]{cases_offline_ess_mse}
  \caption{Mean square error in probability \eqref{eq:mse_prob}.} \label{fig:cases_mse}
  \end{subfigure}
  \caption{Figure \ref{fig:cases_offline_ess} shows the effective sample size and Figure \ref{fig:cases_mse} shows the mean square error in probability (see \eqref{eq:mse_prob}) for the offline estimation procedure run with $B=1$ (green, solid), $B=2$ (orange, dashed) and $B=3$ (purple, dot-dashed) look ahead steps. For each figure, left, middle and right correspond to data simulated according to models \ref{item:type1}, \ref{item:type2} and \ref{item:type3}, respectively.} \label{fig:offline_cases_ess_mse}
\end{figure}

\begin{figure}[t!]
  \begin{subfigure}[t]{\textwidth}
    \centering
  \includegraphics[trim={0 6.5cm 0 1cm}, clip, width=\textwidth]{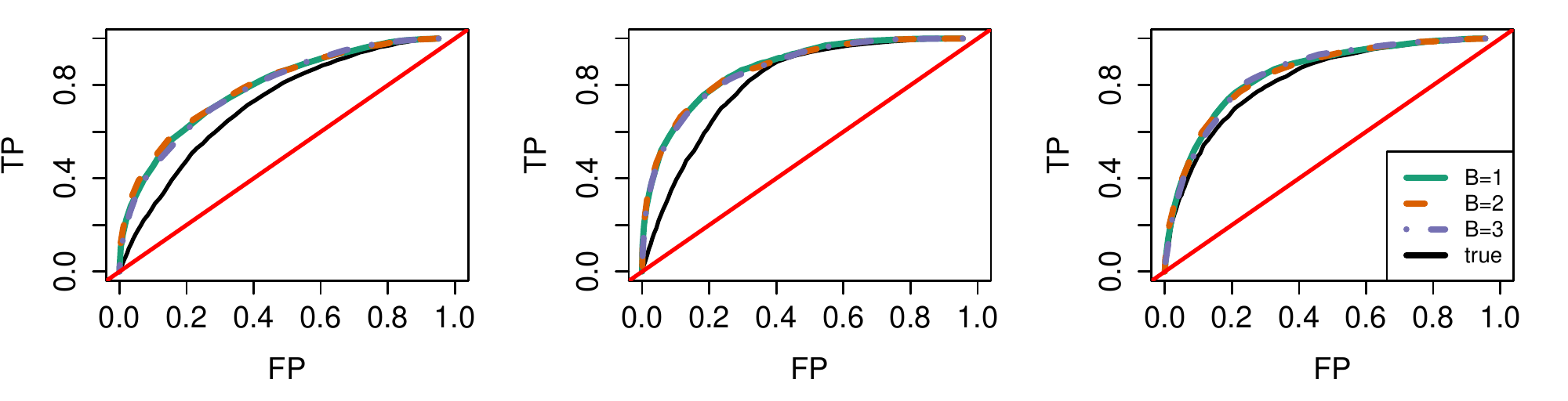}
  \caption{ROC curve for observations at time 1 to time $T-1$.} \label{fig:cases_offline_rocall}
  \end{subfigure}
  \begin{subfigure}[t]{\textwidth}
    \centering
  \includegraphics[trim={0 0cm 0 7cm}, clip, width=\textwidth]{cases_offline_roc}
  \caption{ROC curve for observations at time $T$.} \label{fig:cases_offline_rocpred}
  \end{subfigure}
  \caption{Figures \ref{fig:cases_offline_rocall} and \ref{fig:cases_offline_rocpred} show the ROC curves for observations 1 to $T-1$ and for the predicted probabilities at time $T$, respectively. Each figure reports the ROC for the offline estimation procedure run with $B=1$ (green, solid), $B=2$ (orange, dashed) and $B=3$ (purple, dot-dashed) look ahead steps. For each figure, left, middle and right correspond to data simulated according to models \ref{item:type1}, \ref{item:type2} and \ref{item:type3}, respectively. The line $y=x$ is shown in red and the ROC curve for the true probabilities is shown in black.} \label{fig:offline_cases_rocs}
 \end{figure}

\begin{figure}[h]
  \centering
  \includegraphics[trim={0 0cm 0 1cm}, clip, width=\textwidth]{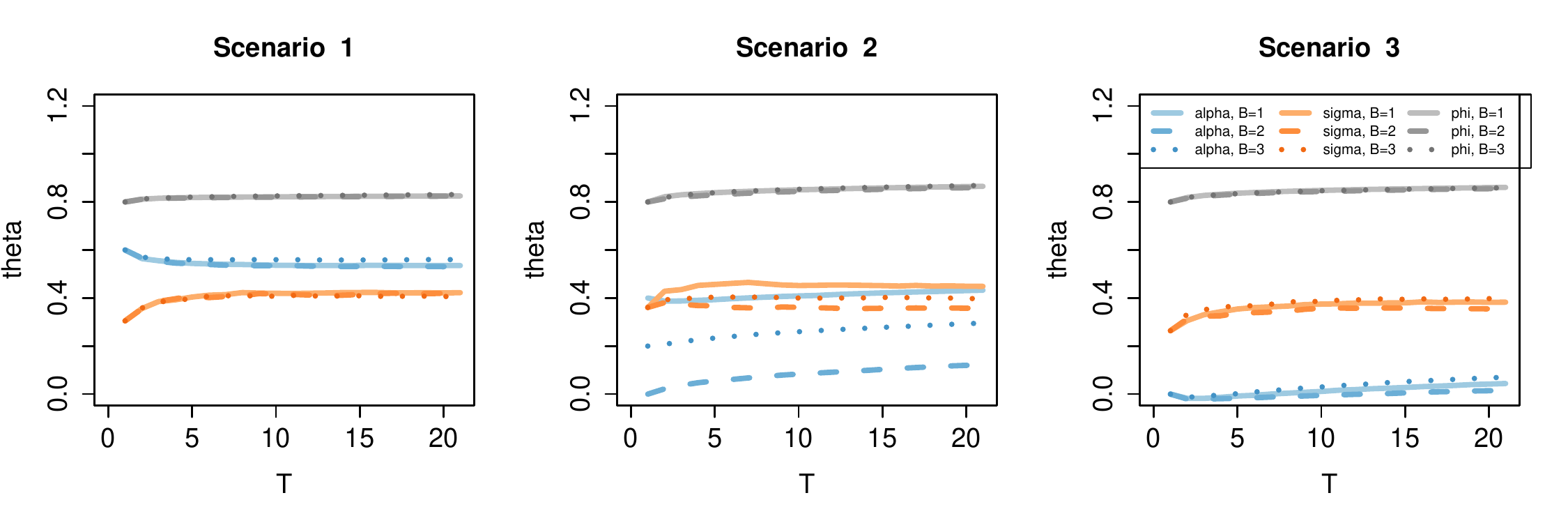}
  \caption{Estimates of $\alpha$ (blue), $\sigma$ (orange) and $\phi$ (grey) obtained via offline procedure. Left, middle and right figures correspond to the data generated according to \ref{item:type1}, \ref{item:type2} and \ref{item:type3}, respectively.} \label{fig:theta_offline_cases}
\end{figure}

Figures \ref{fig:offline_cases_ess_mse}, \ref{fig:offline_cases_rocs} and \ref{fig:theta_offline_cases} summarise the fit for offline estimation procedure for the data scenarios considered in Section \ref{sec:sims_altscen}. Throughout we fix $S=1.5N, M=5000, d=2$ and $N_{\theta}=20$. Corresponding figures for the offline procedure are presented in Section \ref{sec:sims_altscen}.

\section{Performance Plots for Scalability Simulation}
\label{app:scale_perf}

\begin{figure}[h]
  \centering
  \begin{subfigure}[t]{\textwidth}
    \includegraphics[trim={0 6.75cm 0 0}, clip, width=\textwidth]{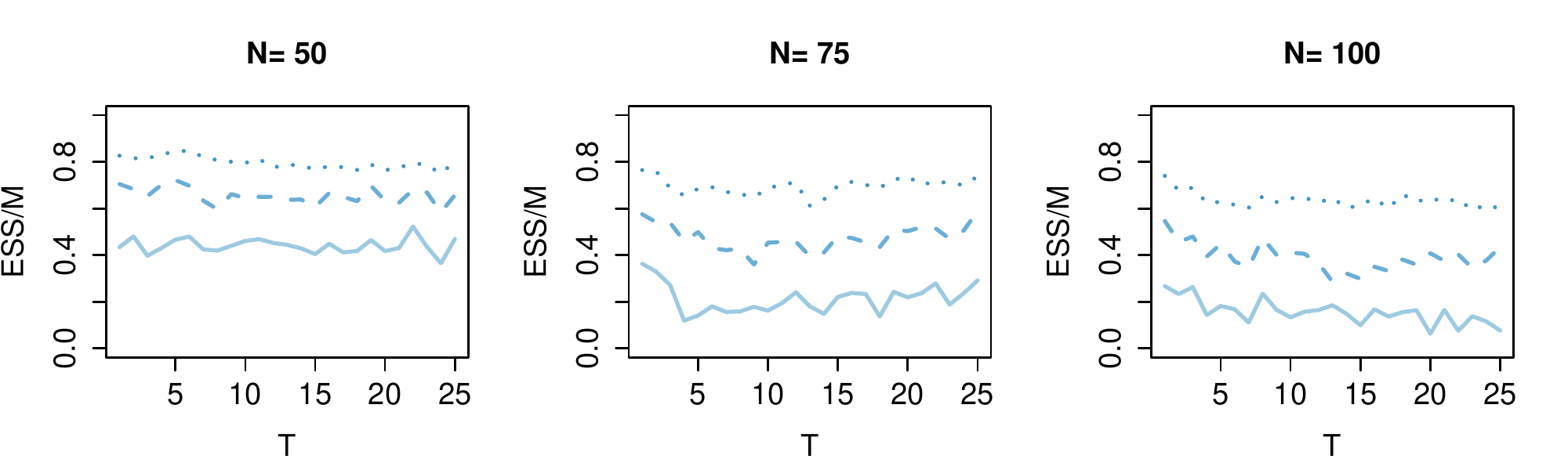}
    \includegraphics[trim={0 0 0 7cm}, clip, width=\textwidth]{scalability_N_essmse}
  \caption{Summary of fit for increasing $N$ with $T=25$ and $S = 0.5N, N,$ and $2N$. The first row shows the ESS scaled by $M$ and the second row shows the MSE in probability. }
\label{fig:scale_summ_Ninc}
  \end{subfigure}
  \begin{subfigure}[t]{\textwidth}
    \centering
  \includegraphics[trim={0 0.5cm 0 1.5cm}, clip, width=.7\textwidth]{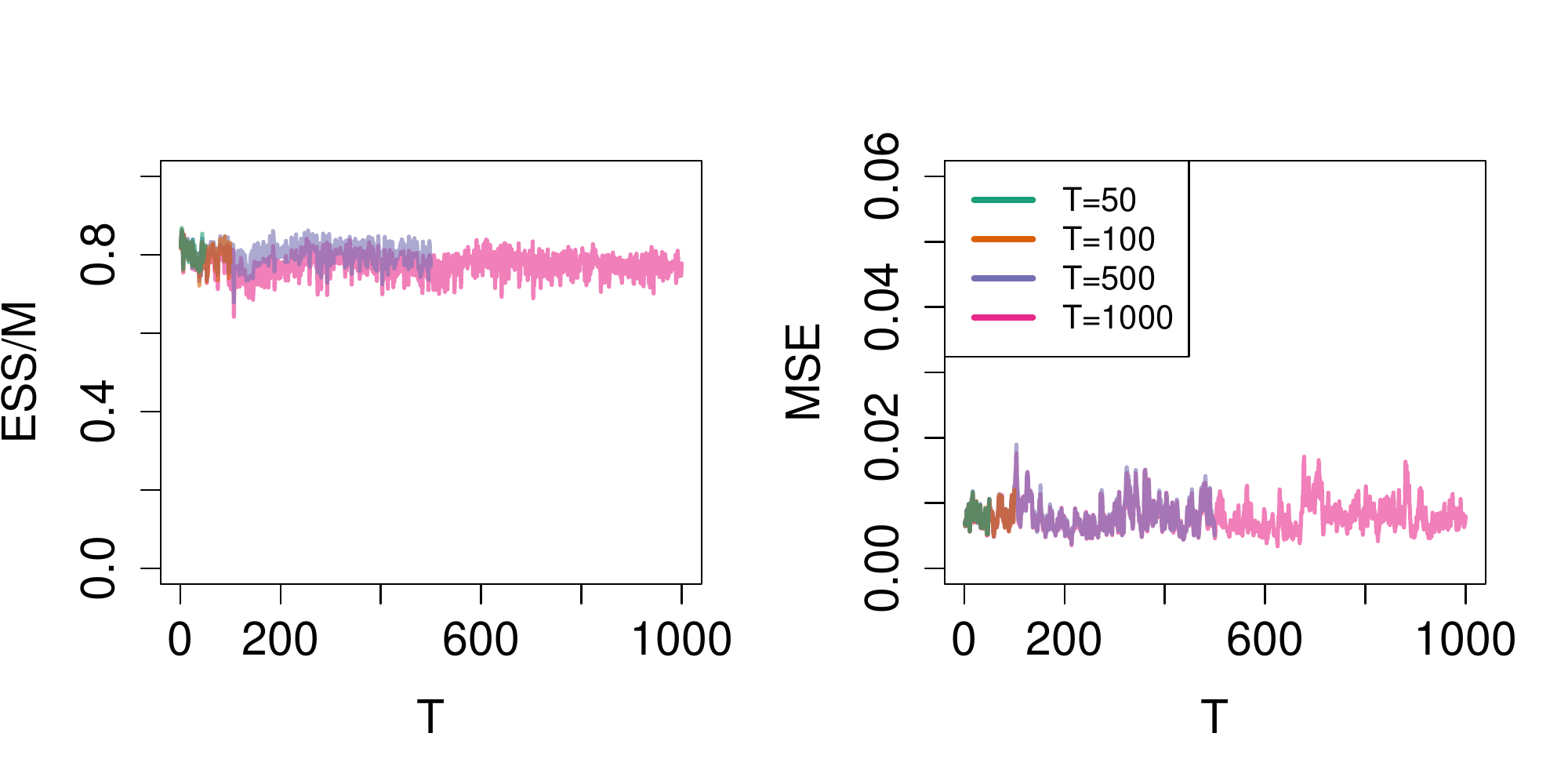}
  \caption{Summary of fit for increasing $T$ with $N=20$ and $S=N$.}
\label{fig:scale_summ_Tinc}
  \end{subfigure}
  \caption{Summary of performance of online SMC scheme as the dimension of the data increases in terms of $N$ and $T$.} \label{fig:scale_summary}
\end{figure}

Figure \ref{fig:scale_summary} shows the performance of the filter as $N$ and $T$ increase and the data were generated according to the description in Section \ref{sec:sims_scalability}. From Figure \ref{fig:scale_summ_Ninc}, we see that the problem becomes more challenging as $N$ grows and choosing $S$ too small will decrease the performance of the filter. From Figure \ref{fig:scale_summ_Tinc} we see that the performance of the filter is stable at $T$ increases. As expected, increasing the dimension of the state space is the more challenging aspect.

\end{document}